\documentclass[11pt,english,aps,superscriptaddress]{revtex4}
\usepackage{amsmath}
\usepackage{graphicx}
\usepackage{subfigure}
\usepackage{multirow}
\usepackage{bm}
\usepackage{amssymb}
\usepackage{mathtools}
\usepackage{enumerate}
\usepackage{ulem}
\usepackage{color}
\usepackage[colorlinks,linkcolor=magenta,anchorcolor=cyan,citecolor=blue]{hyperref}

\begin{document}
\title{Quench Dynamics in Holographic First-Order Phase Transition}

\author{Qian Chen}
\email{chenqian192@mails.ucas.ac.cn}
\affiliation{School of Physical Sciences, University of Chinese Academy of Sciences, Beijing 100049, China}

\author{Yuxuan Liu}
\email{223093@csu.edu.cn}
\affiliation{School of Physics and Electronics, Central South University, Changsha 418003, China}

\author{Yu Tian}
\email{ytian@ucas.ac.cn}
\affiliation{School of Physical Sciences, University of Chinese Academy of Sciences, Beijing 100049, China}
\affiliation{Institute of Theoretical Physics, Chinese Academy of Sciences, Beijing 100190, China}

\author{Xiaoning Wu}
\email{wuxn@amss.ac.cn}
\affiliation{Institute of Mathematics, Chinese Academy of Sciences, Beijing 100190, China}

\author{Hongbao Zhang}
\email{hongbaozhang@bnu.edu.cn}
\affiliation{Department of Physics, Beijing Normal University, Beijing 100875, China}

\begin{abstract}
In this work, we investigate the real-time dynamics of quenching a state from phase separation in a holographic model of first-order phase transition. 
In addition to the typical phase-separated and high-energy final states, we have discovered a novel dynamical process that drives the system to a low-temperature supercooled final state within a narrow range of quench parameters.
The critical behavior is also revealed during the fully non-linear dynamics.
Following a sudden quench with critical parameters, the phase separation can be attracted to a critical nucleus. 
Specifically, the critical nucleus will subsequently shrink in size and eventually disappear for super-critical parameters, where the system is actually supercooled with a temperature lower than the initial one. 
While for sub-critical parameters, the nucleus will grow in size and finally reform a phase separation, where the absorbed quenching energy is reflected in the increment of the latent heat.
\end{abstract}

\maketitle

\section{Introduction}\label{sec:In}
The concept of phase transitions, which signals abrupt changes in the properties of a thermodynamic system across the phase boundary in a phase diagram, is inherently related to equilibrium physics.
Within the framework of equilibrium physics, the real-time dynamics of phase transitions is approximated by the so-called quasi-static process. However, this process, by definition, is too idealized to capture the transition process in real life, which generally goes beyond the realm of thermodynamic and even hydrodynamic descriptions, and enters the far-from-equilibrium regime. 
Thus, it is significant to develop a theoretical framework to describe the real-time dynamics of phase transitions in a more realistic manner. Fortunately, holographic duality provides a desirable framework where a many-body system can be mapped into a few-body gravitational entity with an extra dimension\cite{Maldacena:1997re,Gubser:1998bc,Witten:1998qj,Witten:1998zw}.  
In particular, the real-time dynamics of the many-body system can be extracted from the bulk dynamics of its dual gravitational one, which is always amenable to the numerical simulations even though the analytic treatment is inapplicable\cite{Chesler:2013lia}.

Among others, such a holographic tool has recently demonstrated its unique power in investigating a variety of issues related to the first-order phase transition, from the thermodynamic properties of the equilibrium states\cite{Gubser:2008ny,Dias:2017uyv}, and the near-equilibrium stability by linear response analysis\cite{Janik:2015iry,Janik:2016btb},
all the way to the out-of-equilibrium evolution by fully nonlinear simulations\cite{Gursoy:2016ggq,Attems:2018gou,Attems:2017ezz,Janik:2017ykj,Bellantuono:2019wbn,Attems:2019yqn,Attems:2020qkg,Bea:2020ees,Bea:2021ieq,Bea:2022mfb,Janik:2022wsx,Chen:2022cwi}. 
In particular, the real-time process of the phase separation via both the spinodal decomposition and the nucleation in the supercooled phase is successfully achieved by the bulk gravitational dynamics\cite{Attems:2017ezz,Janik:2017ykj,Bellantuono:2019wbn,Attems:2019yqn,Attems:2020qkg,Bea:2020ees,Chen:2022cwi}. 

Despite the broad interest in phase separation, the ongoing discussion surrounding this phenomenon is predominantly confined to the microcanonical ensemble, where the energy of the system is conserved throughout the evolution.
This limitation restricts our comprehension of the intricate interplay between relevant thermodynamic variables in realistic systems.
The heating process of an ice-water mixture serves as an exemplary instance.
If it is a first-order phase transition, then not only should one find that the heating enlarges the liquid water region and reduce the solid ice region but one also observe that the final coexisting phases are still at the critical temperature. 
Certainly, compared to the aforementioned phase separation dynamics, it is a numerical challenge to realize such a physical process by the dual gravitational simulations because the involved system is essentially an open one rather than the closed one.
The purpose of this paper is to take up this challenge and manage to investigate the real-time dynamics of the heating process based on phase separation.

An additional motivation for this paper is to provide evidence for the nonlinear instability of phase-separated states.
The process of inducing phase separation in a supercooled state involves the formation and growth of seed nuclei.
Such dynamical process exhibits a wealth of critical phenomena \cite{Bea:2020ees,Bea:2022mfb,Chen:2022cwi}, analogs to the type I critical gravitational collapse \cite{Evans:1994pj,Koike:1995jm,Bizon:1998kq,Gundlach:2007gc} and the bald/scalarized black hole transition via a nonlinear accretion of the scalar field \cite{Zhang:2021nnn,Zhang:2022cmu,Liu:2022fxy,Jiang:2023yyn,Chen:2023eru}.
That is to say, there exists a critical nucleus that acts as a dynamical barrier, beyond which nucleation becomes more favorable and the system undergoes phase separation.
A natural question is whether there is a critical dynamics mechanism from the phase-separated state to the supercooled state, analogous to the bidirectional transition between bald black holes and scalarized black holes through a critical state \cite{Zhang:2022cmu}.
The holographic quench mechanism introduced in this paper provides an effective approach for this dynamical process.

The plan of the paper is as follows.
After the introduction in section \ref{sec:In}, we give a brief description of a holographic first-order phase transition model in section \ref{sec:Hs}, where the Ward-Takahashi identity and phase structure are revealed.
In section \ref{sec:Ps}, we review the physical process of the system reaching the phase-separated state from the spinodal region and the supercooled region, respectively.
In section \ref{sec:Hq}, we take the phase-separated state as the initial data and then impose a time-dependent quench on it.
On the one hand, the three characteristic stages that the system undergoes in the quench process are revealed.
On the other hand, the numerical results indicate that the quench mechanism is an effective measure to achieve the dynamical transition from the phase-separated state to the homogeneous state, during which the system exhibits critical behavior.
Finally, we conclude the paper with a summary and an outlook in section \ref{sec:C}.

\section{Holographic setup}\label{sec:Hs}
In this section, we will introduce a holographic model with a first-order phase transition, and then show the Ward-Takahashi identity and phase structure.

\subsection{Holographic model}
We consider the Einstein-Scalar system in the four-dimensional asymptotically anti-de Sitter spacetime described by the Lagrangian density
\begin{equation}
	\mathcal{L}=R-\frac{1}{2}\nabla_{\mu}\phi\nabla^{\mu}\phi-V(\phi)
\end{equation}
with the following scalar potential
\begin{equation}
	V(\phi)=-6\text{cosh}\left(\frac{\phi}{\sqrt{3}}\right)-\frac{\phi^{4}}{5},
\end{equation}
where the AdS radius has been set to unity for convenience.
According to the AdS/CFT correspondence, this gravity system is dual to a boundary conformal field theory with a scalar operator of conformal dimension $\Delta=2$. To explore the operators of the stress tensor and scalar, we need to renormalize the bulk action by adding some boundary terms to make it finite. This can be achieved as follows \cite{Gibbons:1976ue,Bianchi:2001kw,Elvang:2016tzz}:
\begin{equation}
	2\kappa^{2}_{4}S_{\text{ren}}=\int_{M}dx^{4}\sqrt{-g}\mathcal{L}+2\int_{\partial M}dx^{3}\sqrt{-\gamma}K-\int_{\partial M}dx^{3}\sqrt{-\gamma}\left(R[\gamma]+4+\frac{1}{2}\phi^{2}\right),
\end{equation}
where $R[\gamma]$ is the Ricci scalar associated with the induced metric $\gamma_{\mu\nu}$ on the boundary and $K$ is the trace of extrinsic curvature $K_{\mu\nu}=\gamma^{\sigma}_{\mu}\nabla_{\sigma}n_{\nu}$ with $n_{\nu}$ the outward normal vector field to the boundary.
On the one hand, the variation of the action with respect to the bulk fields gives rise to the equations of motion as
\begin{subequations}
	\begin{align}
		G_{\mu\nu}\equiv R_{\mu\nu}-\frac{1}{2}Rg_{\mu\nu}&=\frac{1}{2}\nabla_{\mu}\phi\nabla_{\nu}\phi-\left(\frac{1}{4}\left(\nabla\phi\right)^{2}+\frac{1}{2}V(\phi)\right)g_{\mu\nu},\\
		\nabla^{\mu}\nabla_{\mu}\phi&=\frac{dV(\phi)}{d\phi}.
	\end{align}\label{eq:fe}
\end{subequations}
On the other hand, the aforementioned two dual operators are given by
\begin{subequations}
	\begin{align}
		\left\langle O\right\rangle &=\kappa_{4}^{2}\lim_{r\rightarrow \infty}\frac{r^{2}}{\sqrt{-\gamma}}\frac{\delta S_{\text{ren}}}{\delta{\phi}}=-\frac{1}{2}\lim_{r\rightarrow \infty}r^{2}\left(\phi+ n^{\mu}\nabla_{\mu}\phi\right),\\
		\left\langle T_{ij}\right\rangle&=-2\kappa_{4}^{2}\lim_{r\rightarrow \infty}\frac{r}{\sqrt{-\gamma}}\frac{\delta S_{\text{ren}}}{\delta\gamma^{ij}}=\lim_{r\rightarrow \infty}r\left[G[\gamma]_{ij}-K_{ij}-\left(2-K+\frac{1}{4}\phi^{2}\right)\gamma_{ij}\right].
	\end{align}\label{eq:operators}
\end{subequations}
Accordingly, the variation of the renormalized on-shell action can be expressed as
\begin{equation}
	\kappa^{2}_{4}\delta S_{\text{ren}}=\int_{\partial M}d^{3}x\sqrt{-\gamma_{(0)}} \left(-\frac{1}{2}\left\langle T_{ij}\right\rangle\delta\gamma_{(0)}^{ij}+\left\langle O\right\rangle\delta\phi_{(0)}\right),
\end{equation}
where the subscripts ``$(0)$'' denote the coefficients of the leading order term in the asymptotic behavior of the corresponding bulk fields near the boundary.

Such a renormalized on-shell action is invariant under the following diffeomorphism
\begin{equation}
	\delta \gamma^{ij}=\pounds_{\xi}\gamma^{ij},
	\quad \delta\phi=\pounds_{\xi}\phi,
\end{equation}
where $\pounds_{\xi}$ is the Lie derivative with respect to an arbitrary vector field $\xi^{i}$ tangent to the boundary. Whence one can derive the Ward-Takahashi identity 
\begin{equation}
	\nabla^{j}\left\langle T_{ij}\right\rangle=\left\langle O\right\rangle \nabla_{i}\phi_{(0)}.\label{eq:WD}
\end{equation} 
In particular, following the coordinate system introduced in Appendix~\ref{sec:Ab}, the time component of the above equations reduces to 
\begin{equation}
	\partial_{t}\epsilon-\partial_{x}J_{x}-\partial_yJ_y=-\left\langle O\right\rangle\partial_{t}\phi_{(0)},\label{eq:energy_change}
\end{equation}
where $\epsilon$ denotes the energy density, $J_{x}$ and $J_y$ represent the momentum density in the spatial directions $x$ and $y$, respectively. In what follows, we shall focus on the case in which the system is homogeneous along the $y$-direction and periodic along the $x$-direction. Therefore, the temporal evolution of the total energy of the system is determined by how the scalar operator responds to the time-dependent scalar source. 

\subsection{Phase diagram}

\begin{figure}
	\begin{center}
		\subfigure[]{\includegraphics[width=.49\linewidth]{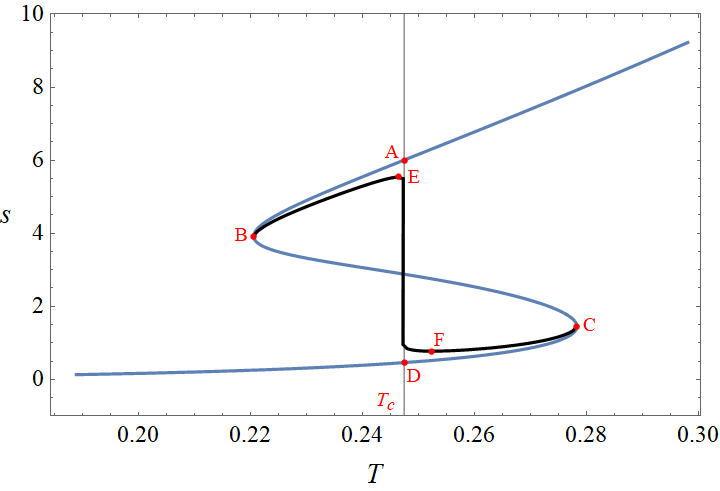}\label{fig:s-T}}
		\subfigure[]{\includegraphics[width=.49\linewidth]{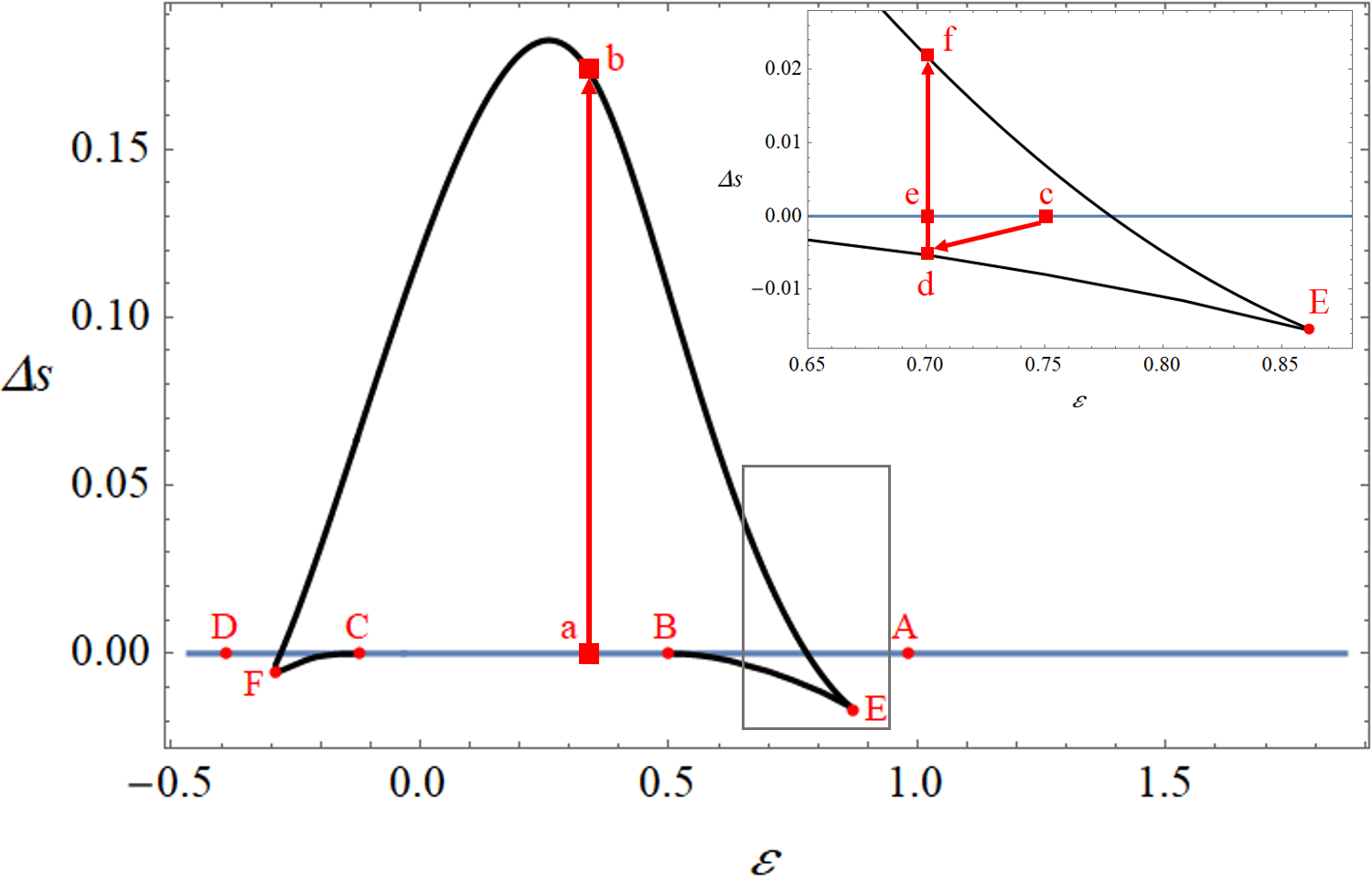}\label{fig:deltas-epsilon}}
		\caption{Canonical phase diagram (a): the entropy density as a function of the temperature.
			Microcanonical phase diagram (b): the entropy density difference between the inhomogeneous and homogeneous states as a function of the energy density.
			The blue and black lines represent homogeneous and inhomogeneous states, respectively.
			The vertical gray line represents the phase transition temperature.
			The schematic paths $(a,b)$ and $(c,d,f)$ correspond to the dynamical processes shown in the left and right panels of Fig. \ref{fig:phase_separation}, respectively.}\label{fig:phase_diagram}
	\end{center}
\end{figure}

In order to reveal the phase structure of this model, we obtain the static solutions to the field equations (\ref{eq:fe}) numerically using both the Deturk method \cite{Dias:2015nua} and the nonlinear evolution method \cite{Chesler:2013lia}, which are described in Appendices~\ref{sec:Aa} and \ref{sec:Ab}, respectively. 
The resulting phase diagram is displayed in Fig. \ref{fig:phase_diagram}, where the blue and black lines represent homogeneous and inhomogeneous phases, respectively. 

According to the first law of thermodynamics $F=-sdT$ with $F$ being the free energy density, the homogeneous phases can be classified into three types. 
The first type is the thermodynamically stable phase, which lies on the upper branch with an entropy density greater than that of point $A$ and on the lower branch with an entropy density less than that of point $D$. 
The positions of points $A$ and $D$ are determined by requiring that the corresponding free energy densities are equal to each other at the critical temperature $T_c=0.247$.
The difference in the entropy density between points $A$ and $D$ indicates the discontinuity of the first derivative of the free energy density with respect to the temperature at $T_c$, signaling a thermal first-order phase transition in the canonical ensemble. 
The second type is referred to as the metastable phase, which includes the supercooled region $AB$ and the superheated region $CD$. 
The third type is the spinodal region between points $B$ and $C$, which is thermodynamically unstable. 
Interestingly, there is a close relationship between thermodynamic instability and dynamical instability. 
Roughly speaking, the thermodynamically stable (unstable) phase is also linearly dynamically stable (unstable) while the metastable phase is linearly dynamically stable but nonlinearly dynamically unstable. 

The intermediate dynamical process of the above nonlinearly dynamical instability as well as the final state driven by either linearly or nonlinearly dynamical instability is related to the inhomogeneous phase in Fig. \ref{fig:phase_diagram}. 
Specifically, the inhomogeneous state in the region $BE$ and the region $CF$ with only one linearly unstable mode serve as the critical state for the metastable phase to undergo toward the linearly stable phase-separated phase lying in the region $EF$. 
Such a phase-separated phase, formed by phase A and phase D joined by a domain wall in between, also serves as the final state of the thermodynamically unstable phase under tiny perturbations. 
{Since the dual boundary system of the inhomogeneous state has spatial dependence, the physical results here are affected by the size of the boundary space, which is reflected in the fact that the black curve representing the inhomogeneous states in the figure adjusts accordingly as the size of the system changes.
On the one hand, in a finite volume, the spatial size can limit the spinodal instability, causing points $B$ and $C$ to move toward the center of the spinodal region \cite{Bea:2020ees}.
On the other hand, in the infinite volume limit, points $E$ and $F$ will coincide with points $A$ and $D$ respectively, resulting in the entropy density of the phase separation region covering the entire $AD$ interval.
In our work, we choose the period length in $x$-direction of space to be a finite but large value $L_{x}=24\pi$ to suppress the influence of the size effect on dynamics to a certain extent.}

To facilitate a clear discussion of the two distinct types of dynamical transition mechanisms involving linearly and nonlinearly dynamical instabilities mentioned above, we present the microcanonical phase diagram in Fig. \ref{fig:deltas-epsilon} as an energy density dependence of the entropy density difference between the inhomogeneous and homogeneous states.
In the microcanonical ensemble, the system tends to reside in the state of maximum entropy.
The presence of spinodal instability in the $BC$ region implies that the system, at the state located at point $a$, will undergo spontaneous evolution toward the phase-separated state at point $b$, exhibiting greater entropy but the same energy.
In contrast, the states belonging to $AB$ and $CD$ regions manifest a distinct behavior. 
Without loss of generality, we consider the state at point $c$ as an example, where the system remains stable under small perturbations.
To initiate the dynamical transition, a disturbance of sufficient strength is required as a seed nucleus.
Furthermore, if the disturbance strength approaches a threshold, the system will display a critical behavior, which proceeds through three stages. 
First, the addition of the seed nucleus leads to a sudden reduction in the energy and entropy of the system. 
Subsequently, the system swiftly converges to the critical state at point $d$. 
Finally, the phase-separated state with maximum entropy at point $f$ is reached for supercritical strength, and conversely, the system evolves to the homogeneous state with local maximum entropy at point $e$.
The details of the real-time dynamics during the dynamical transitions are presented in Fig. \ref{fig:phase_separation}, which we shall describe in the next section.

\section{Phase separation}\label{sec:Ps}
\begin{figure}
	\begin{center}
		\subfigure[]{\includegraphics[width=.49\linewidth]{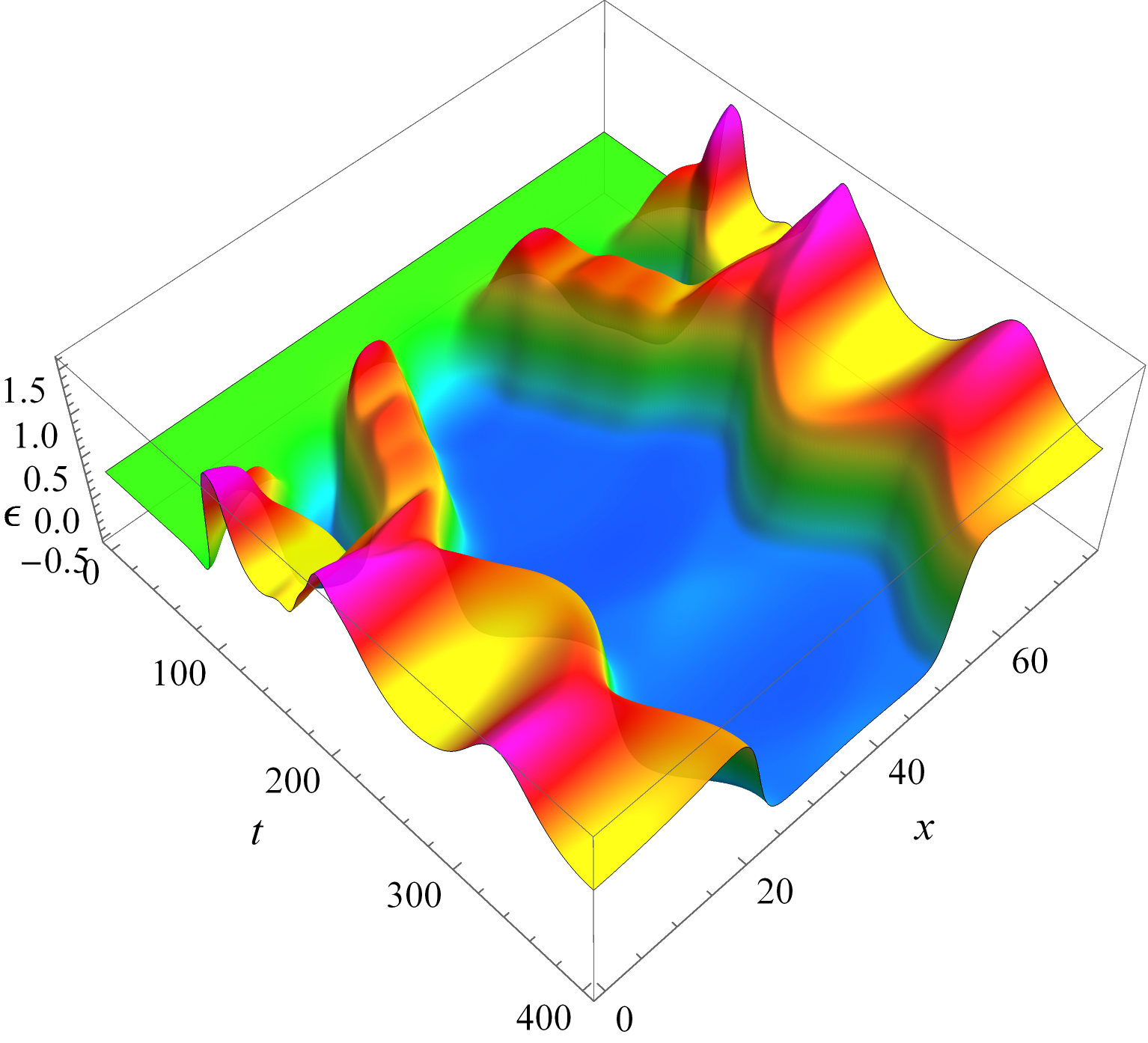}\label{fig:ps_spinodal_energy}}
		\subfigure[]{\includegraphics[width=.49\linewidth]{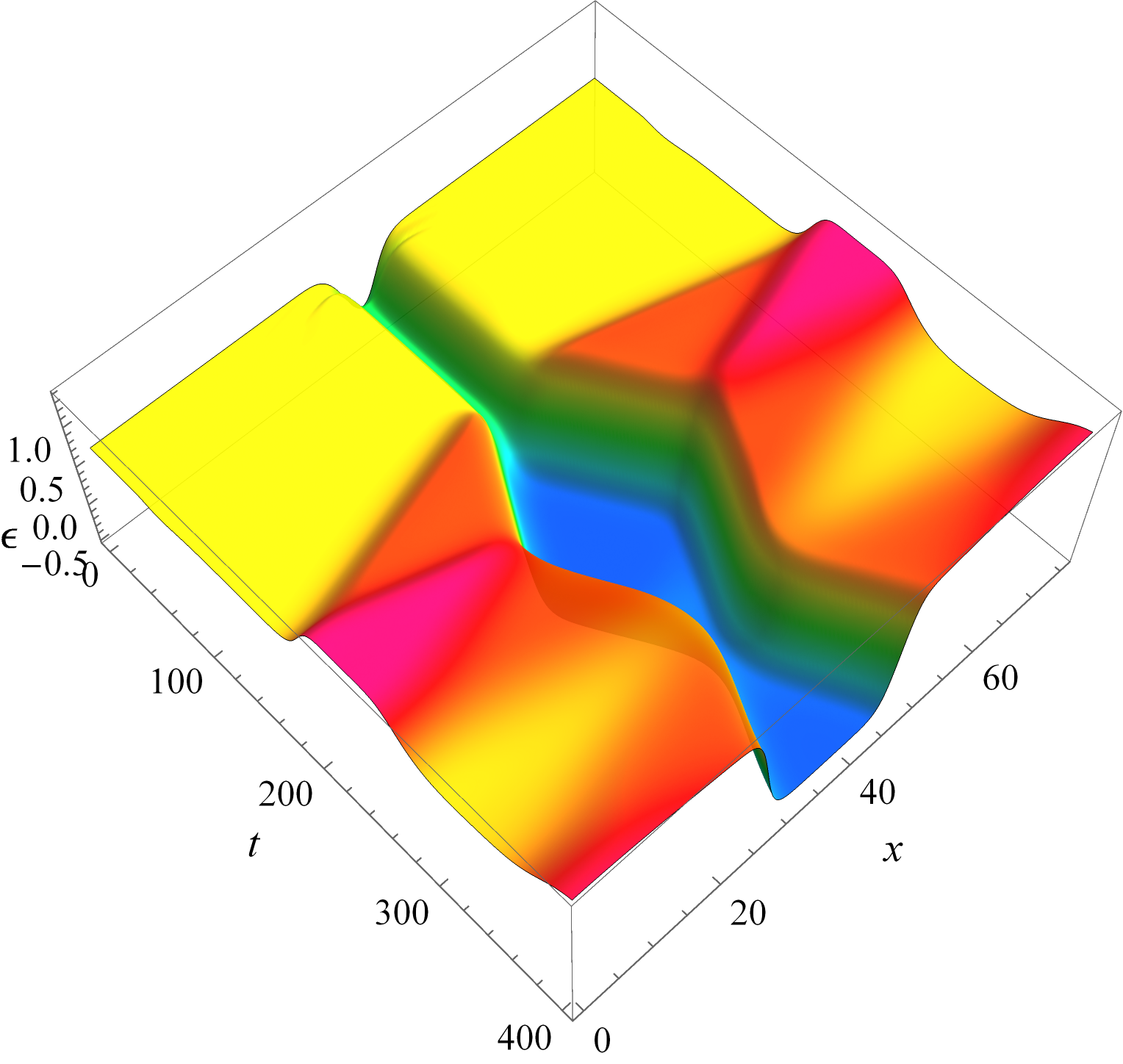}\label{fig:ps_supercool_energy}}
		\subfigure[]{\includegraphics[width=.49\linewidth]{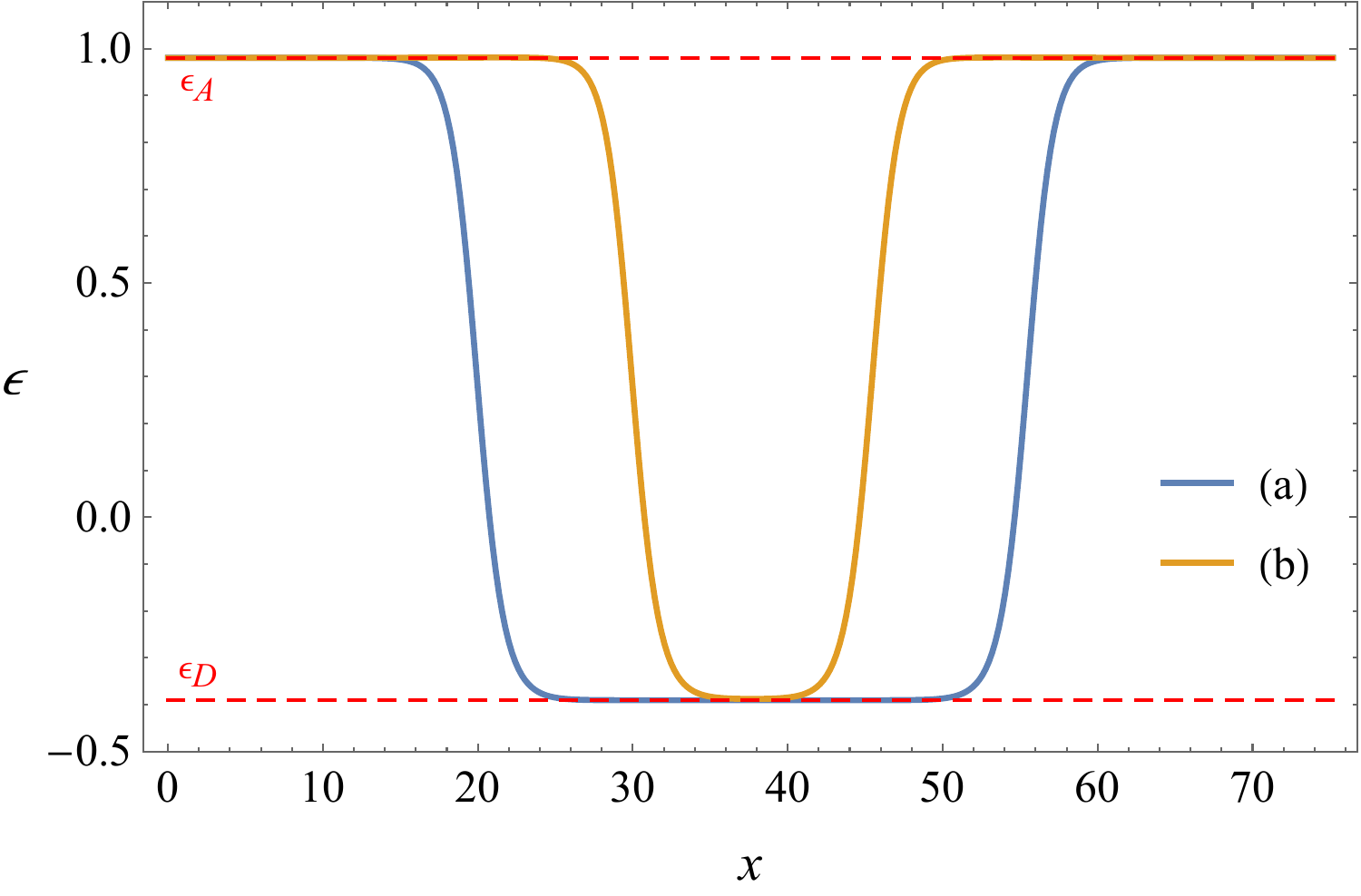}\label{fig:energy_final}}
		\subfigure[]{\includegraphics[width=.49\linewidth]{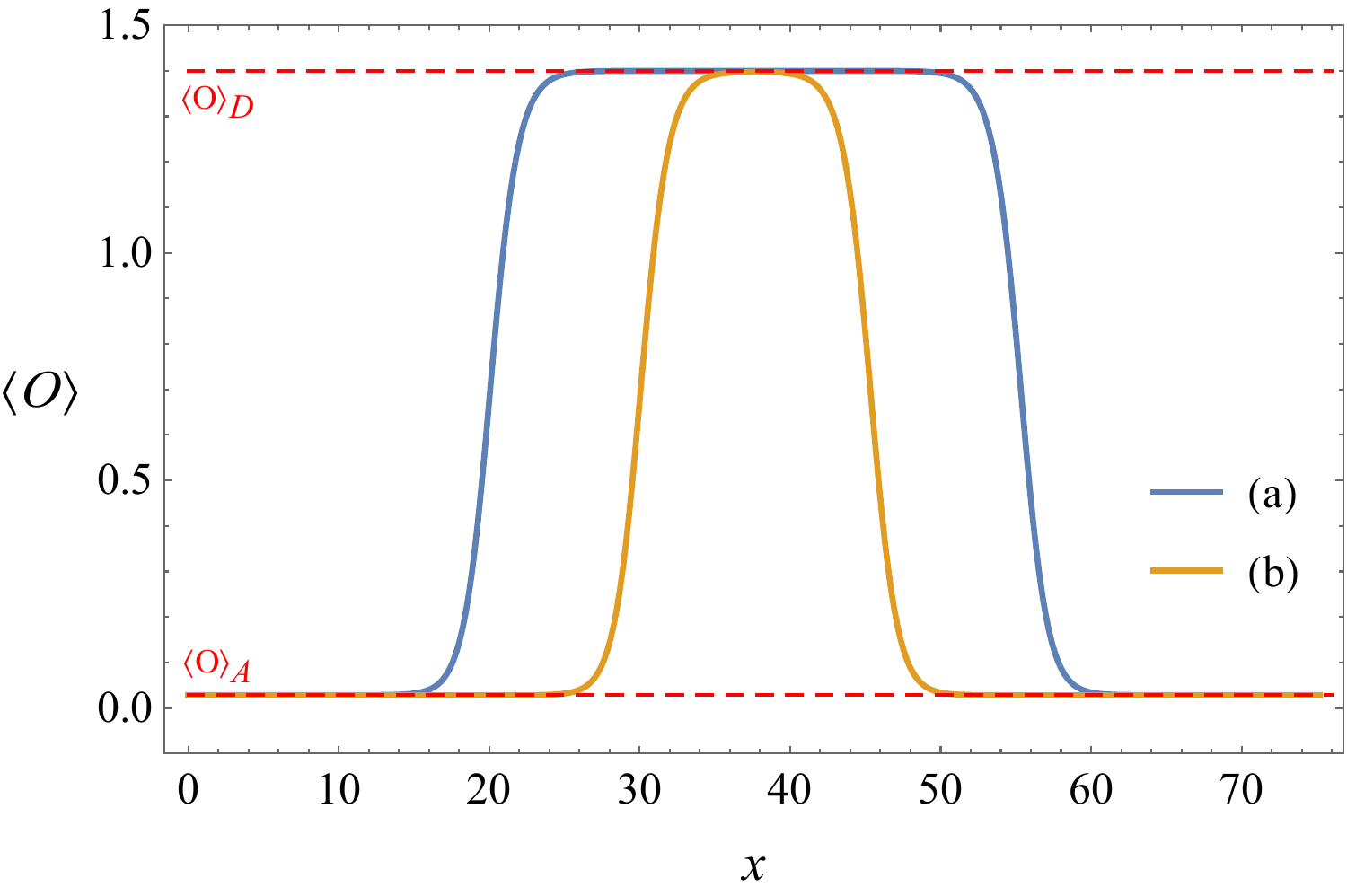}\label{fig:scalar_final}}
		\caption{(a, b): The energy density as a function of time with the initial states at points $a$ in the spinodal region and $c$ in the supercooled region in Fig. \ref{fig:deltas-epsilon}, respectively.
			(c, d): The spatial dependence of the energy density and expectation value of the scalar field of the final state, {which is extracted from the dynamical evolution at time $t=3000$}.
			The blue and orange lines correspond to the final states given in processes (a) and (b), respectively.
			The horizontal red dotted lines represent the states at the phase transition temperature in homogeneous solutions (points $A$ and $D$ in Fig. \ref{fig:s-T}).
			Note that the high-energy phase has a lower expected value of the scalar field, while the low-energy phase possesses a higher one.}\label{fig:phase_separation}
	\end{center}
\end{figure}

In this section, we perform fully nonlinear numerical simulations of the dual gravitational dynamics to generate the phase-separated state, which serves as the initial data for the subsequent heating process. 
A detailed description of the numerical procedure can be found in Appendix \ref{sec:Ab}, and the source code is available at \href{https://github.com/QianChen2022/HFOPT}{https://github.com/QianChen2022/HFOPT}.
To clarify the two aforementioned dynamical transition mechanisms, we select the initial states at points $a$ and $c$ in the spinodal and supercooled regions of Fig. \ref{fig:deltas-epsilon}, respectively.
The resulting real-time dynamics are depicted in Fig. \ref{fig:ps_spinodal_energy} and Fig. \ref{fig:ps_supercool_energy}, respectively.

In the case of spontaneous dynamical transition, the initial state may undergo spinodal instability, where even tiny perturbations can trigger a dynamical transition in the system. To account for this, we introduce an $x$-dependent perturbation to the scalar field, given by the following expression
\begin{equation}
	\delta\phi=-0.1z^{2}(1-z)^{2}\text{exp}\left[-10\text{cos}^{2}\left(x/24\right)\right].\label{eq:10}
\end{equation}
{The $z^{2}$ term in the form of perturbation depends on the asymptotic behavior of the scalar field near the AdS boundary (\ref{eq:asy_phi}), where the coefficient of the linear term $\phi_{1}$ is fixed to $1$ in the dynamical process.
Since the initial position of the horizon is at $z=1$, the existence of the term $(1-z)^{2}$ indicates that the perturbation converges to $0$ at the horizon.}
As can be seen from Fig. \ref{fig:ps_spinodal_energy}, this form of perturbation excites three energy peaks on the background of the homogeneous solution. Notably, due to the periodic boundary condition, the peaks on both ends of the space are identical.
Subsequently, two of these energy peaks slowly migrate towards each other and eventually merge to form a larger phase, leading to a phase-separated state as shown in Figs. \ref{fig:energy_final} and \ref{fig:scalar_final}. Specifically, this state consists of the phases at points $A$ and $D$ in Fig. \ref{fig:deltas-epsilon}, connected by domain walls.

In contrast, when the dynamical transition originates from the supercooled region, the absence of unstable modes in the initial state prevents small perturbations from destabilizing the system.
To initiate the dynamical transition, we introduce a perturbation with a large enough amplitude to overcome the dynamical barrier, serving as a seed nucleus for the system. The expression for this perturbation is given by
\begin{equation}
	\delta\phi=pz^{2}\text{exp}\left[-50\text{cot}^{2}\left(x/24\right)\right],\label{eq:3.2}
\end{equation}
{where the perturbation amplitude $p=1.285364$ is chosen to be slightly larger than the threshold $p_{*}$ required for the dynamical transition.
The value of the threshold $p_{*}$ depends on the specific form of perturbation.
We find that perturbations to the horizon can effectively reduce the value of the threshold.
Therefore, compared to the form (\ref{eq:10}), the term $(1-z)^{2}$ is absent here}.
As shown in Fig. \ref{fig:ps_supercool_energy}, the disturbed system quickly converges to a critical state at early times and remains in this state for a prolonged period. Eventually, the nucleus grows and leads to the formation of the low-energy phase, culminating in the phase-separated state.
As the initial energy of the system, in this case, is higher than that of the spontaneous dynamical transition, the ratio of space occupied by the high-energy phase to that occupied by the low-energy phase in the final state is naturally greater than that observed in the spontaneous dynamical transition, as depicted in Figs. \ref{fig:energy_final} and \ref{fig:scalar_final}.

Remarkably, the system displays a critical behavior similar to that observed in gravitational collapse and scalarization phenomena during this process.
Specifically, there exists a threshold $p_{*}$ at which the system just stays in the critical state.
For supercritical parameter $p>p_{*}$, the system undergoes a dynamical transition, and otherwise, the system returns to a homogeneous state.
For $p$ near $p_{*}$, the system is attracted to the critical state and remains in this state for a duration satisfying
\begin{equation}
	\tau\propto-\gamma\text{ln}\left(|p-p_{*}|\right),
\end{equation}
where the critical exponent $\gamma$ is equal to the reciprocal of the eigenvalue of the only unstable eigenmode of the critical state.
For more details, please refer to \cite{Chen:2022cwi}.

\section{Holographic quench}\label{sec:Hq}
In this section, we expand upon the analysis presented in the previous section by investigating the nonlinear dynamics of a time-dependent scalar source on the background solution of the phase-separated state.
It is worth noting that the confining AdS boundary effectively restricts matter from escaping, which implies that the two dynamical transition processes discussed earlier occur under the microcanonical ensemble. Consequently, the total energy of the system remains conserved during the evolution.
However, as indicated by the Ward-Takahashi identity (\ref{eq:WD}), the time-dependent scalar source can induce changes in the total energy, expressed as
\begin{equation}
	\partial_{t}E=-\int dx\left\langle O\right\rangle\partial_{t}\phi_{(0)}.\label{eq:pt_E}
\end{equation}
It is important to note that periodic boundary conditions are imposed in the spatial $x$-direction. Furthermore, in addition to the energy, the electric charge of a charged system also undergoes changes under the time-dependent scalar source boundary conditions \cite{Bhaseen:2012gg,Bai:2014tla,Chen:2022vag}.

\subsection{Quench dynamics}\label{sec:Dq}

\begin{figure}
	\begin{center}
		\subfigure[]{\includegraphics[width=.49\linewidth]{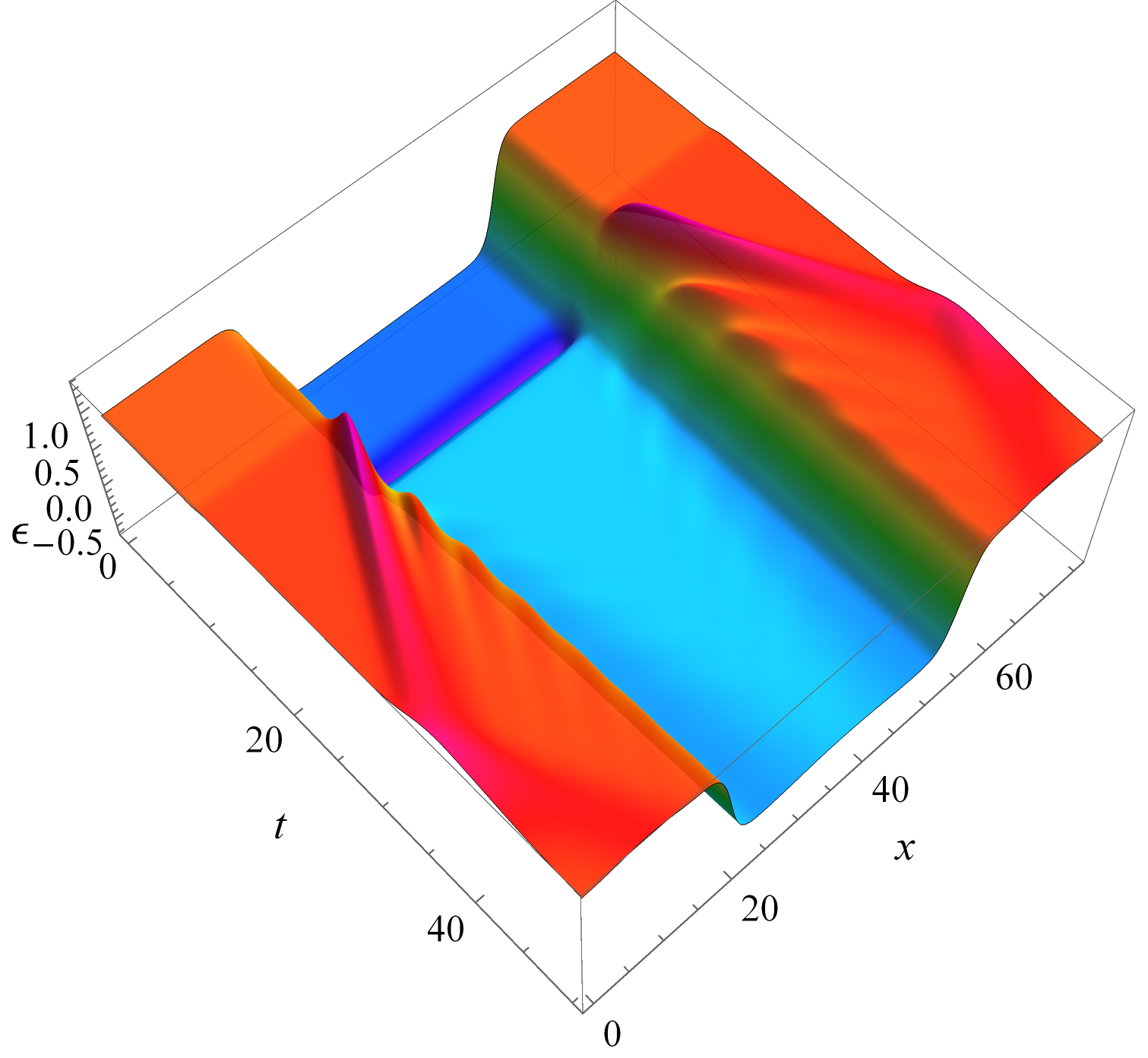}\label{fig:energy_density_quench}}
		\subfigure[]{\includegraphics[width=.49\linewidth]{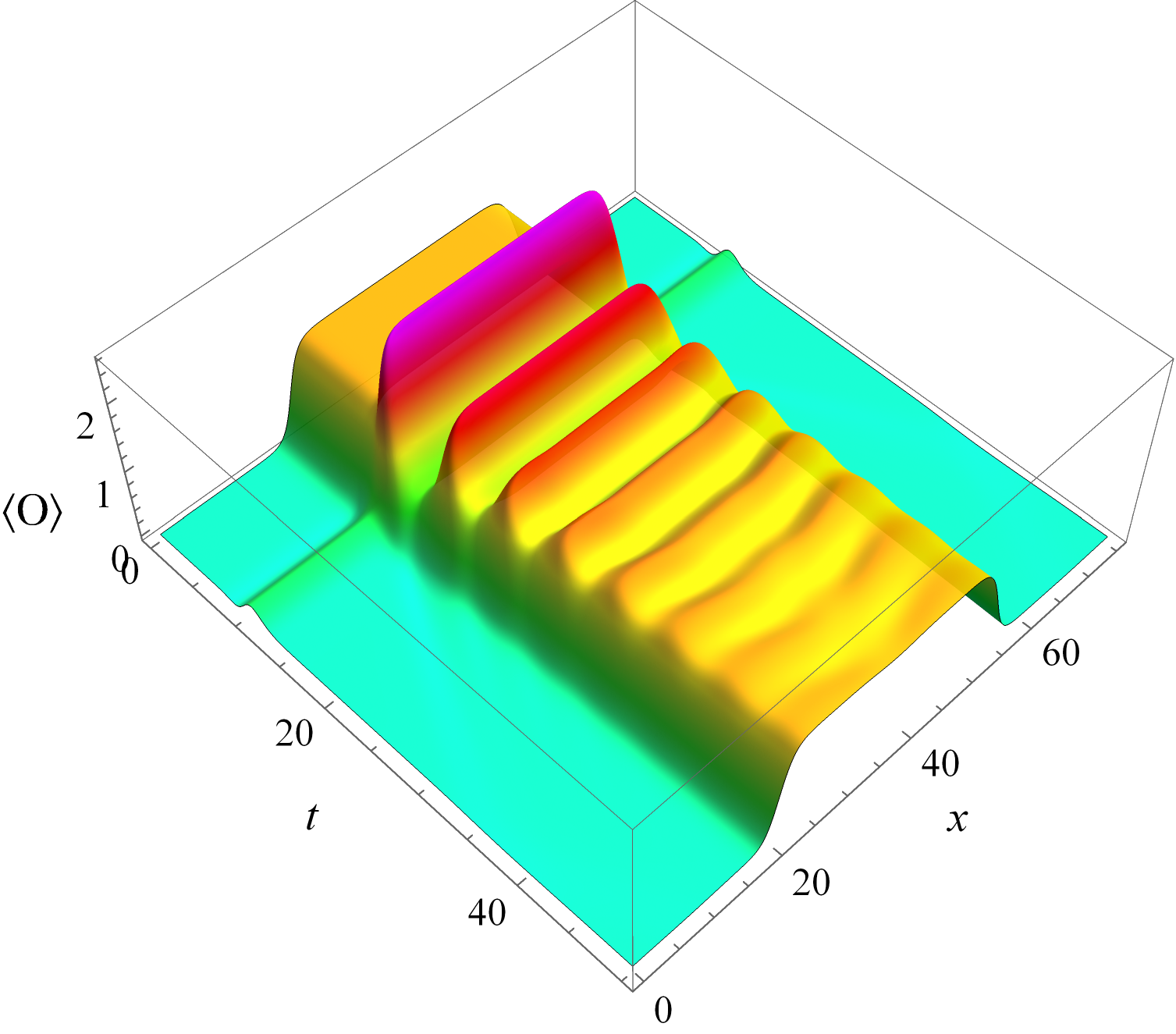}\label{fig:scalar_quench}}
		\caption{The temporal and spatial dependence of the energy density (a) and the expectation value of scalar field (b) in the quench process with quench strength $H=0.2$.}\label{fig:real_time_quench}
	\end{center}
\end{figure}

\begin{figure}
	\begin{center}
		\subfigure[]{\includegraphics[width=.49\linewidth]{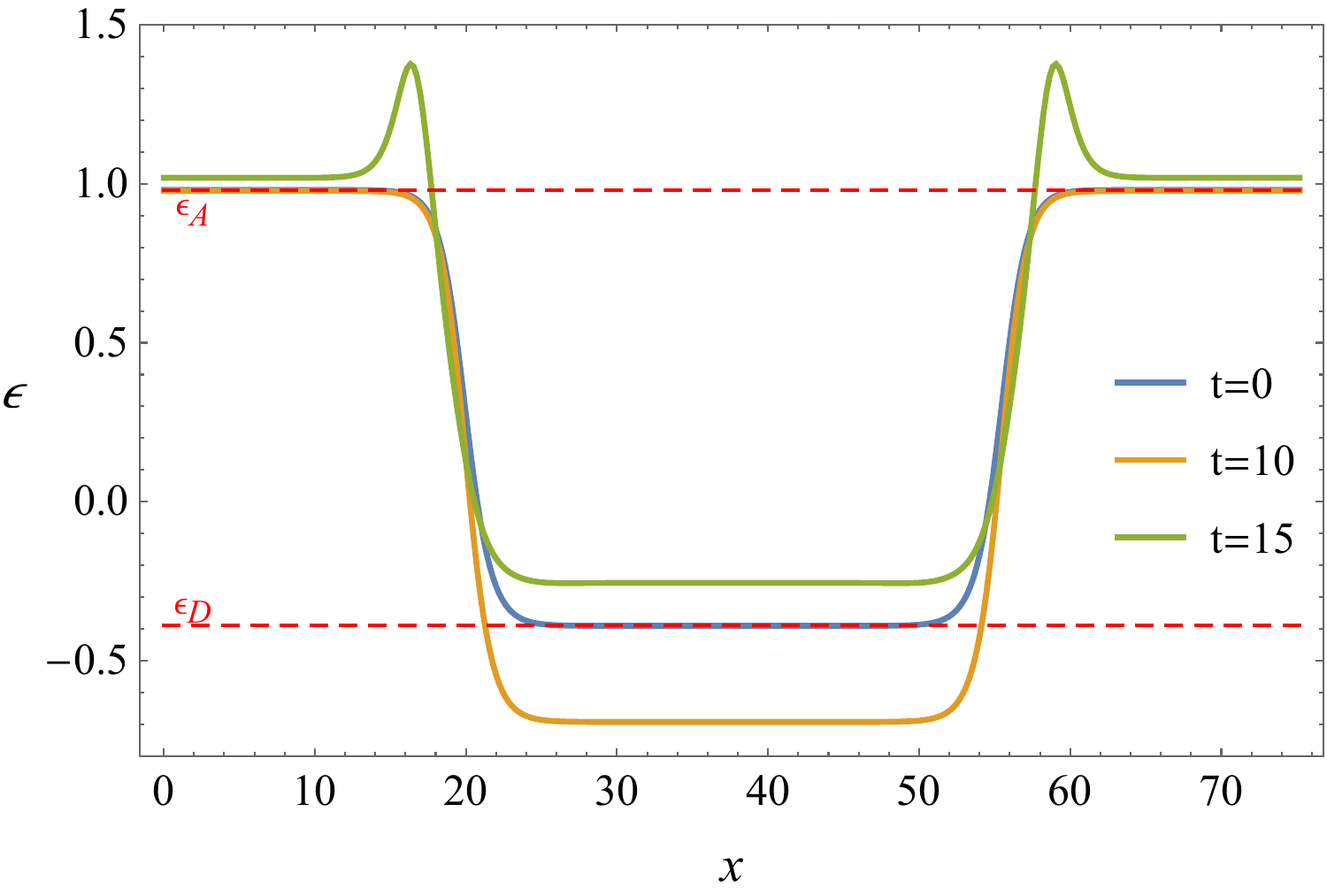}\label{fig:stage1}}
		\subfigure[]{\includegraphics[width=.49\linewidth]{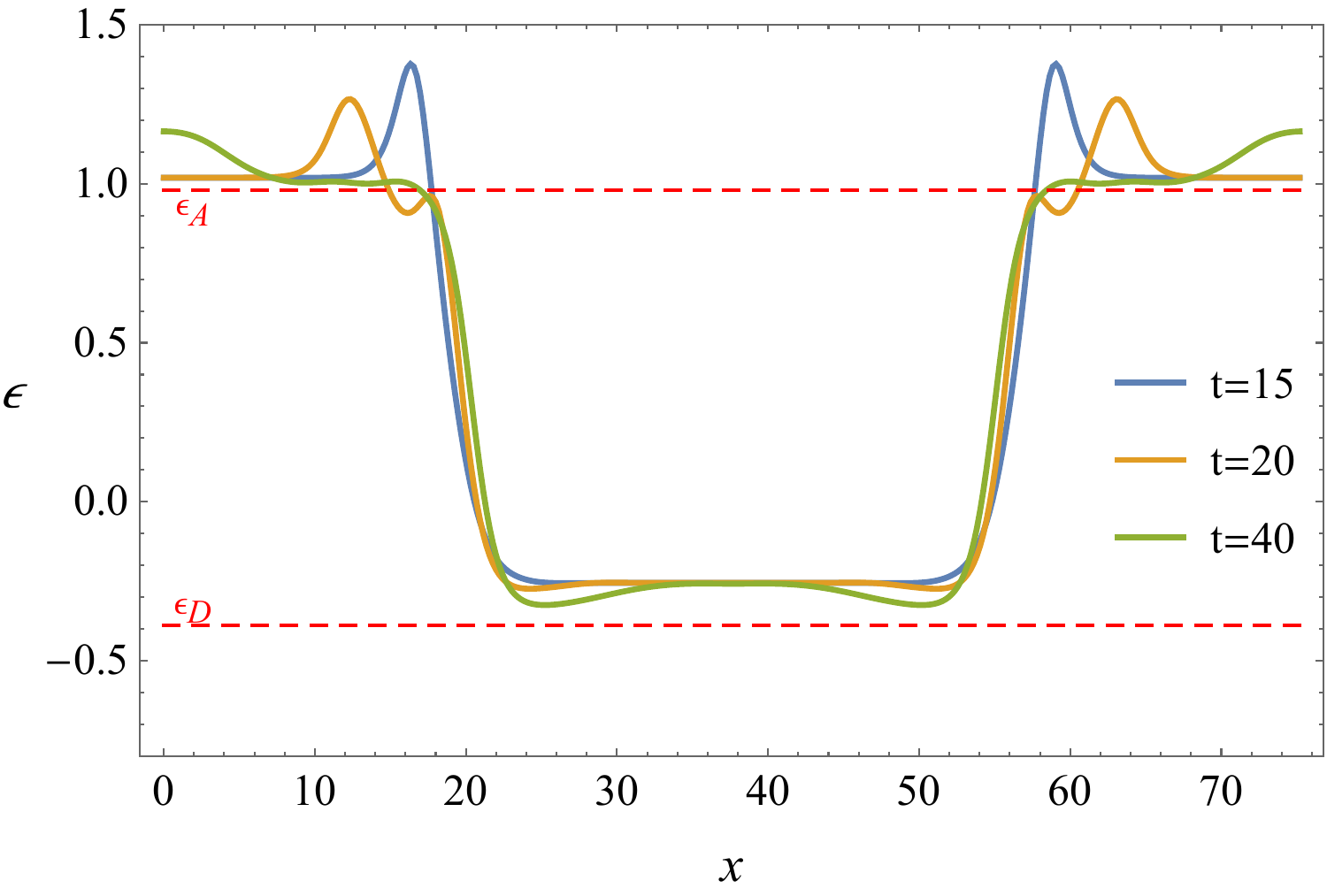}\label{fig:stage2}}
		\subfigure[]{\includegraphics[width=.49\linewidth]{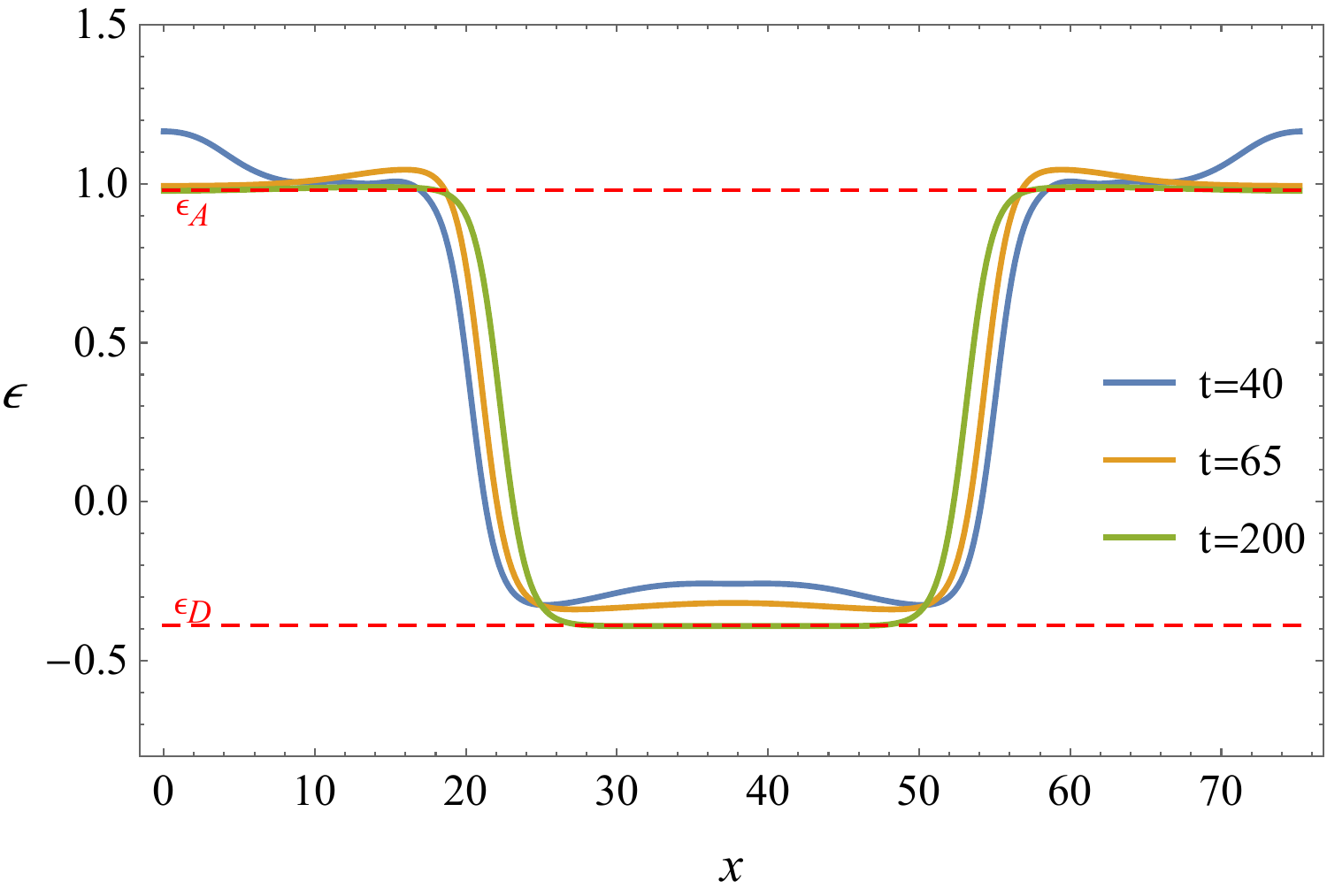}\label{fig:stage3}}
		\subfigure[]{\includegraphics[width=.49\linewidth]{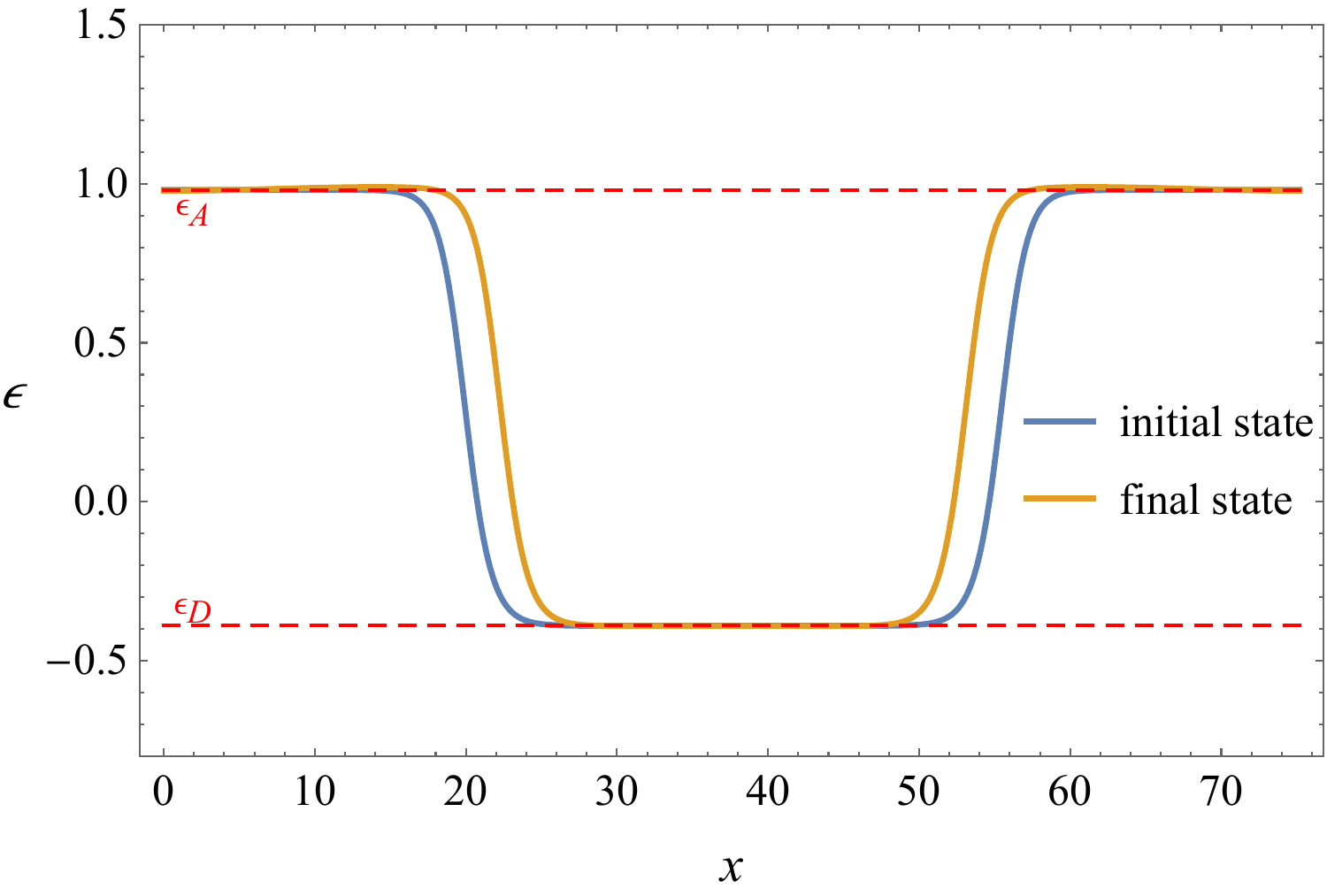}\label{fig:initial_final}}
		\caption{(a,b,c): Three characteristic stages in the quench process. The solid lines of different colors represent the energy density of the system at different times.
			The horizontal red dotted lines represent the energy density of the states at the phase transition temperature in homogeneous solutions.
			(d): The energy density of the initial and final states during the quench process.
			{Among them, the final state is extracted from the dynamical evolution at time $t=3000$.}}\label{fig:stages_quench}
	\end{center}
\end{figure}

To investigate the system's response to the scalar source, we impose a time-dependent behavior on the scalar source over the phase-separated state obtained from the process depicted in Fig. \ref{fig:ps_spinodal_energy}, given by
\begin{equation}
	\phi_{(0)}=\Lambda+H\text{exp}\left[-0.5\left(t-10\right)^{2}\right],
\end{equation}
where the parameter $H$ denotes the strength of the quench, while the source rapidly decays to the constant $\Lambda$ at late times, which serves as the boundary condition for the dynamical transition processes. In our study, we set $\Lambda$ to be equal to $1$.
The real-time dynamics during the quench process with a strength of $H=0.2$ are illustrated in Fig. \ref{fig:real_time_quench}.

\begin{figure}
	\begin{center}
		\subfigure[]{\includegraphics[width=.49\linewidth]{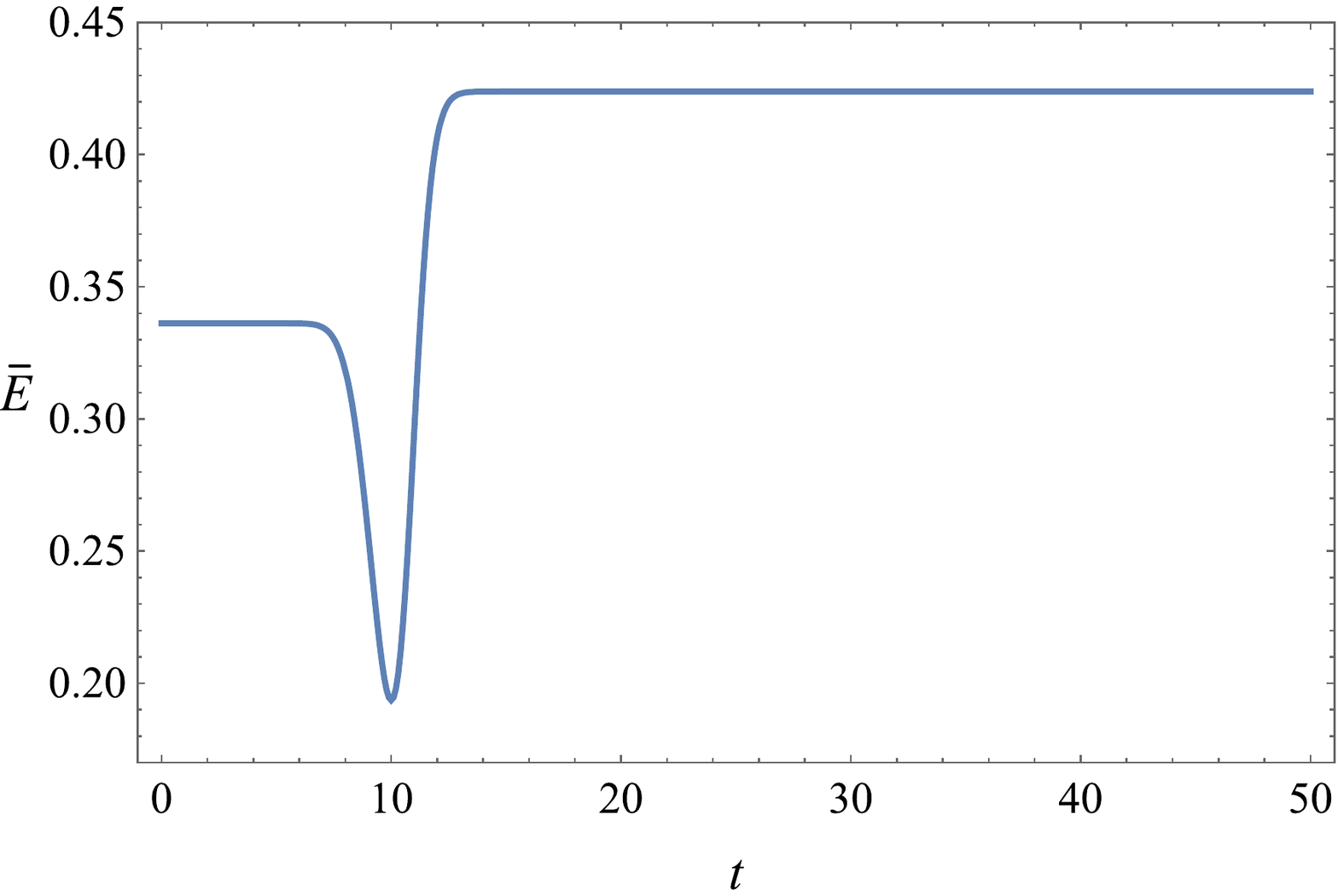}\label{fig:energy_T}}
		\subfigure[]{\includegraphics[width=.49\linewidth]{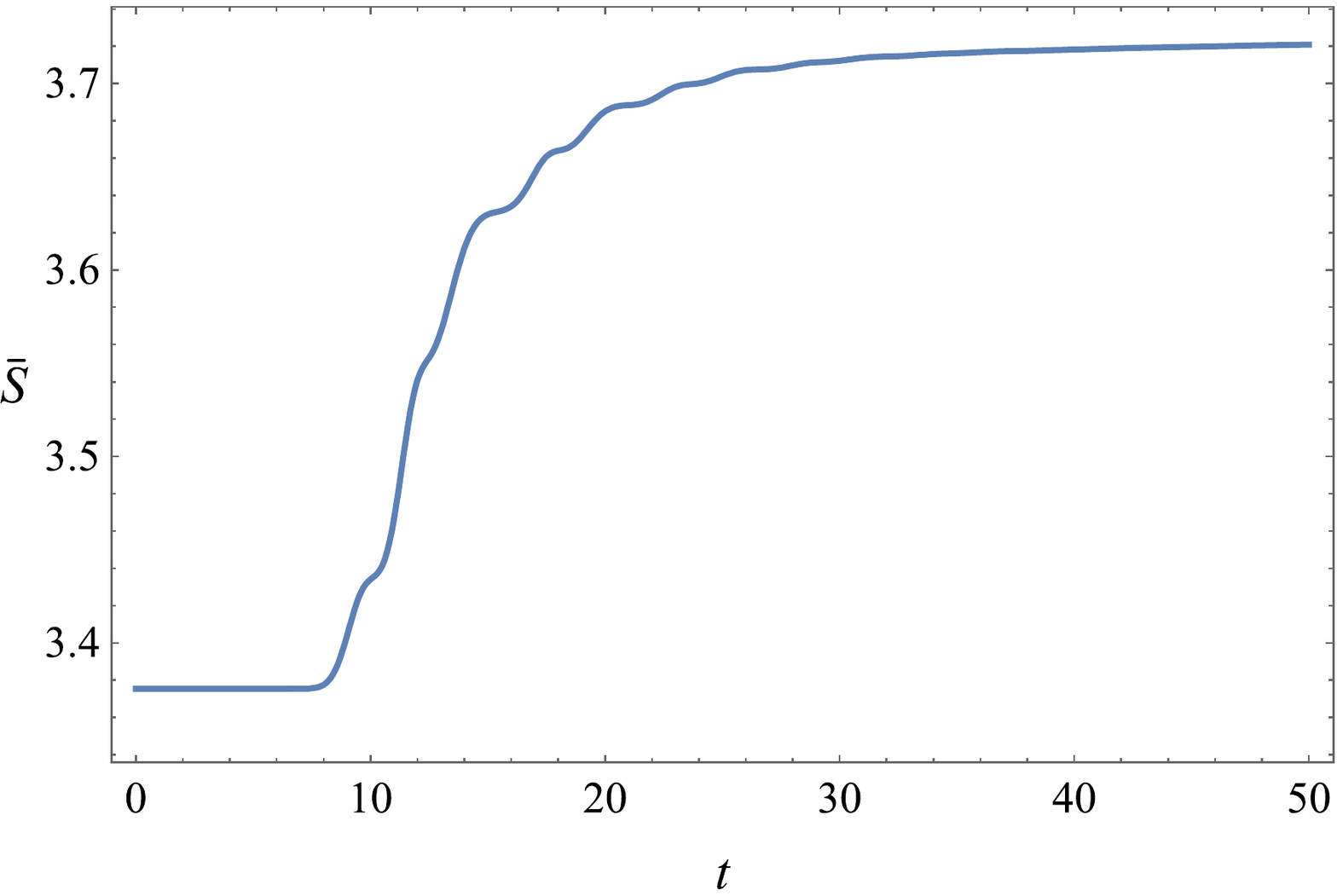}\label{fig:entropy_T}}
		\caption{The average energy (a) and the average entropy (b) of the system as a function of time during quenching with $H=0.2$.}\label{fig:physics_T}
	\end{center}
\end{figure}

The overall evolution process can be divided into three distinct stages. During the first stage, the energy density variation is primarily influenced by the combination of the time-dependent scalar source and the scalar field's expectation value.
Notably, there is no energy flow in the space region occupied by the two phases, resulting in the phenomenon where the energy density can only shift up and down, as depicted in Fig. \ref{fig:stage1}. This stage corresponds to a process in which the two phases are each independently quenched.
Since the low-energy phase possesses a larger expectation value, its energy density is more sensitive to the scalar source, which can be easily observed from (\ref{eq:energy_change}).
Specifically, when $t<10$, the energy density of the low-energy phase decreases in tandem with the increasing source due to its positive expected value. Conversely, the energy density of the high-energy phase remains relatively unchanged during this period, owing to its expected value being close to zero.
When $10<t<15$, as the source decays back to its initial value $\Lambda$, the energy density curves of both phases rise and stabilize at a new location where the energy density is higher than the initial value.
That is to say, after the quench, the energy of both phases increases.
This stage is characterized by a variation in the total energy of the system with the scalar source.
After this, the total energy is conserved since the scalar source is a constant, as shown in Fig. \ref{fig:energy_T}.
At the end of the first stage, an intriguing phenomenon occurs with the behavior of the domain walls.
Two energy peaks are excited at the junctions of domain walls and the high-energy phase.
And in the second stage shown in Fig. \ref{fig:stage2}, these energy peaks propagate towards the interior of the high-energy phase.
Eventually, at late times, they merge together at the center of the phase domain after colliding.
At the time $t=20$, the second set of energy peaks is excited again at the domain walls, which move outwards and are absorbed by the first set of energy peaks subsequently.
During this stage, multiple groups of energy peaks propagating outward are excited successively at the domain walls, which are shown as sawtooth-like ripples in Fig. \ref{fig:energy_density_quench}.
They end up merging together to form a larger one at time $t=40$.
{Finally, due to the periodic boundary conditions, the energy peaks pass through the edges of the system and are converted into inward energy waves in the third stage shown in Fig. \ref{fig:stage3}.}
They then hit the domain walls with the result that part of them goes through and the other part bounces off.
As the energy waves repeatedly propagate through space and gradually decay, the system eventually converges to a phase-separated state with the same properties as the initial state but with a larger total energy shown in Fig. \ref{fig:initial_final}.
During the quench process, the expectation value of the scalar field predominantly exhibits oscillatory behavior, as depicted in Fig. \ref{fig:scalar_quench}.

The time dependence of the average energy and average entropy defined as $\bar{S}=\frac{2\pi}{L_{x}}\int dx\Sigma^{2}\left(r_{h}\right)$ in the quench process are shown in Fig. \ref{fig:physics_T}, where $\Sigma^{2}\left(r_{h}\right)$ represents the area of the apparent horizon and $L_{x}$ stands for the length of the box.
Since the integral of the expectation value of the scalar field is always positive, the change in energy follows the opposite trend of the scalar source, as depicted in Fig. \ref{fig:energy_T}.
Although the energy can increase or decrease during the quench process, the final state always has a higher energy than the initial state.
The reason for this is that, on the one hand, the changes in the physical quantities of the initial and final states satisfy the following first law of black hole thermodynamics
\begin{equation}
	\Delta E=T_{c}\Delta S.\label{eq:first_law}
\end{equation}
Note that $T_{c}$ is the phase transition temperature, which is also the temperature of the initial and final states during the quench process. 
And on the other hand, the second law of black hole mechanics requires the entropy of the system never decreases in dynamical processes, which is shown in Fig. \ref{fig:entropy_T}.
Therefore, the right-hand side of the above equation is always positive, indicating that the energy of the system cannot be extracted through the quench process.
This behavior differs from that of charged systems \cite{Bhaseen:2012gg,Bai:2014tla,Chen:2022vag}.

\begin{figure}
	\begin{center}
		\subfigure[]{\includegraphics[width=.49\linewidth]{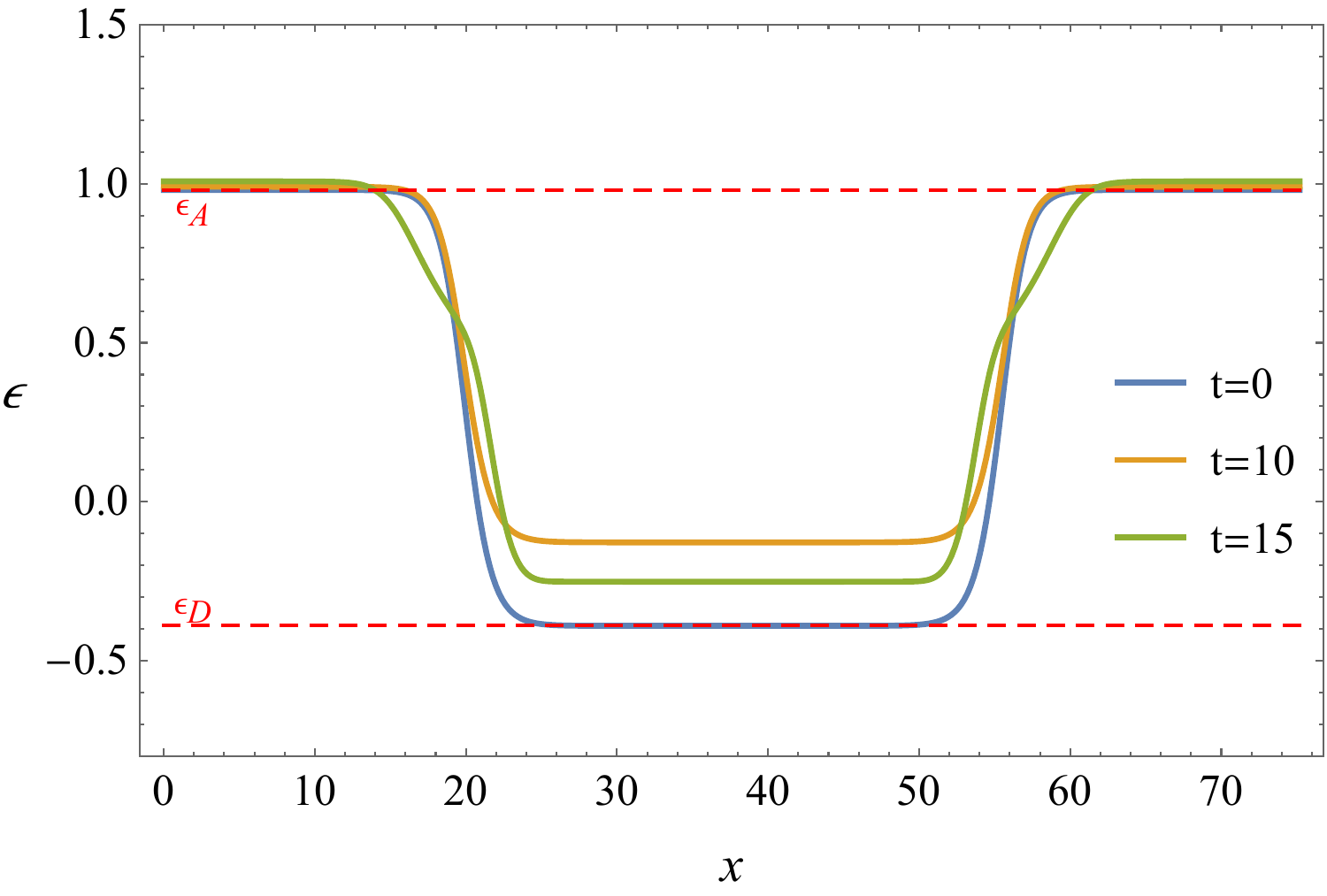}\label{fig:stage1_negative}}
		\subfigure[]{\includegraphics[width=.49\linewidth]{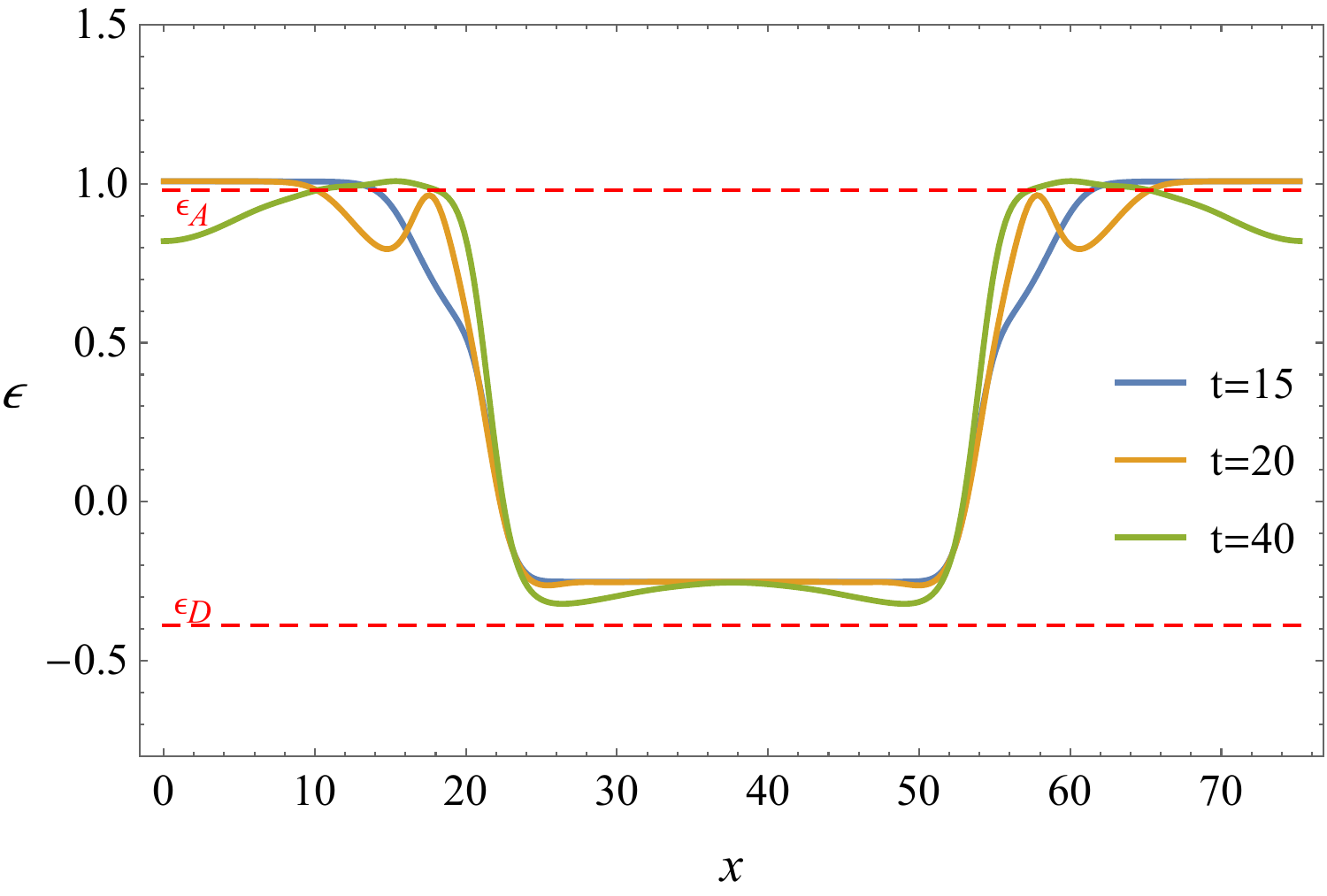}\label{fig:stage2_negative}}
		\caption{The first (a) and second (b) stages in the quench process with quench strength $H=-0.2$.}\label{fig:stages_negative_quench}
	\end{center}
\end{figure}

There is a similar but distinct phenomenon for the case where quench strength $H$ is negative. 
The first and second stages of this dynamic process with $H=-0.2$ are illustrated in Fig. \ref{fig:stages_negative_quench}.
Since the third stage is the stage where the system is in the linear region of the phase-separated state, similar to the process in Fig. \ref{fig:stage3}, we will only focus on the dynamical behavior of the system in the first two stages here.
As depicted in Fig. \ref{fig:stage1_negative}, during the first stage, the energy density in the two-phase region oscillates similarly to that in Fig. \ref{fig:stage1}, as a response to the activation and deactivation of the scalar source.
Since the monotonicity of the scalar source variation is opposite to the case of positive $H$, the up and down motion of the energy curve is also reversed.
Of particular note is the dynamical behavior of the domain walls, which differs fundamentally from the previous case. 
No energy peaks are excited in this scenario, instead, the domain walls undergo significant changes in position. 
The configuration of domain walls is broken at some point in the middle, and then the top half expands while the bottom half contracts.
As shown in Fig. \ref{fig:stage2_negative}, in the second stage, domain walls in the upper half gradually evolve into a set of energy valleys propagating outward.
Finally, they collide in the middle of the region occupied by the high-energy phase and merge together.
Actually, at this stage, multiple groups of energy valleys are generated from the domain walls and are absorbed one by one, similar to the phenomenon in Fig. \ref{fig:stage2}.

\begin{figure}
	\begin{center}
		\subfigure[]{\includegraphics[width=.49\linewidth]{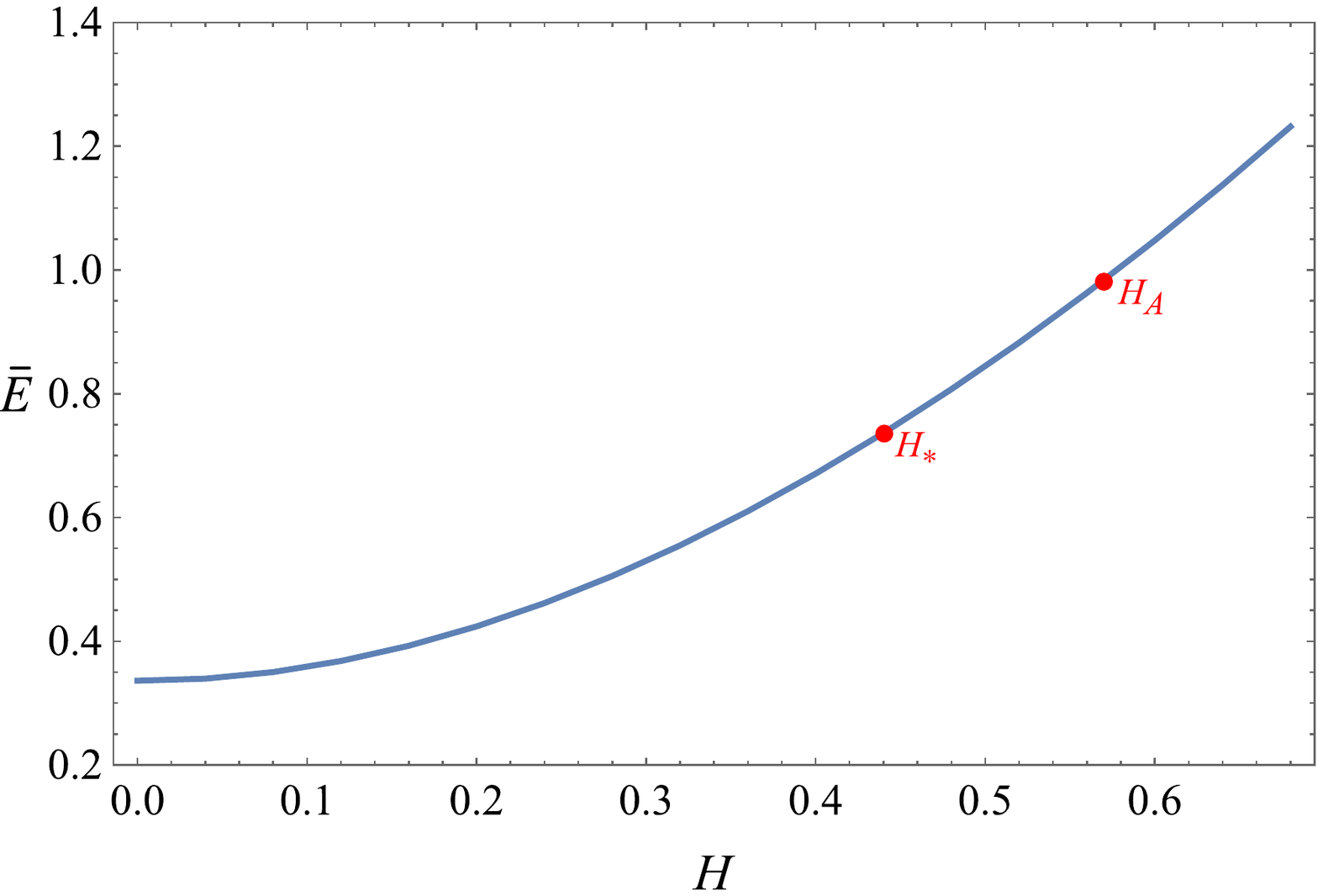}\label{fig:energy_H}}
		\subfigure[]{\includegraphics[width=.49\linewidth]{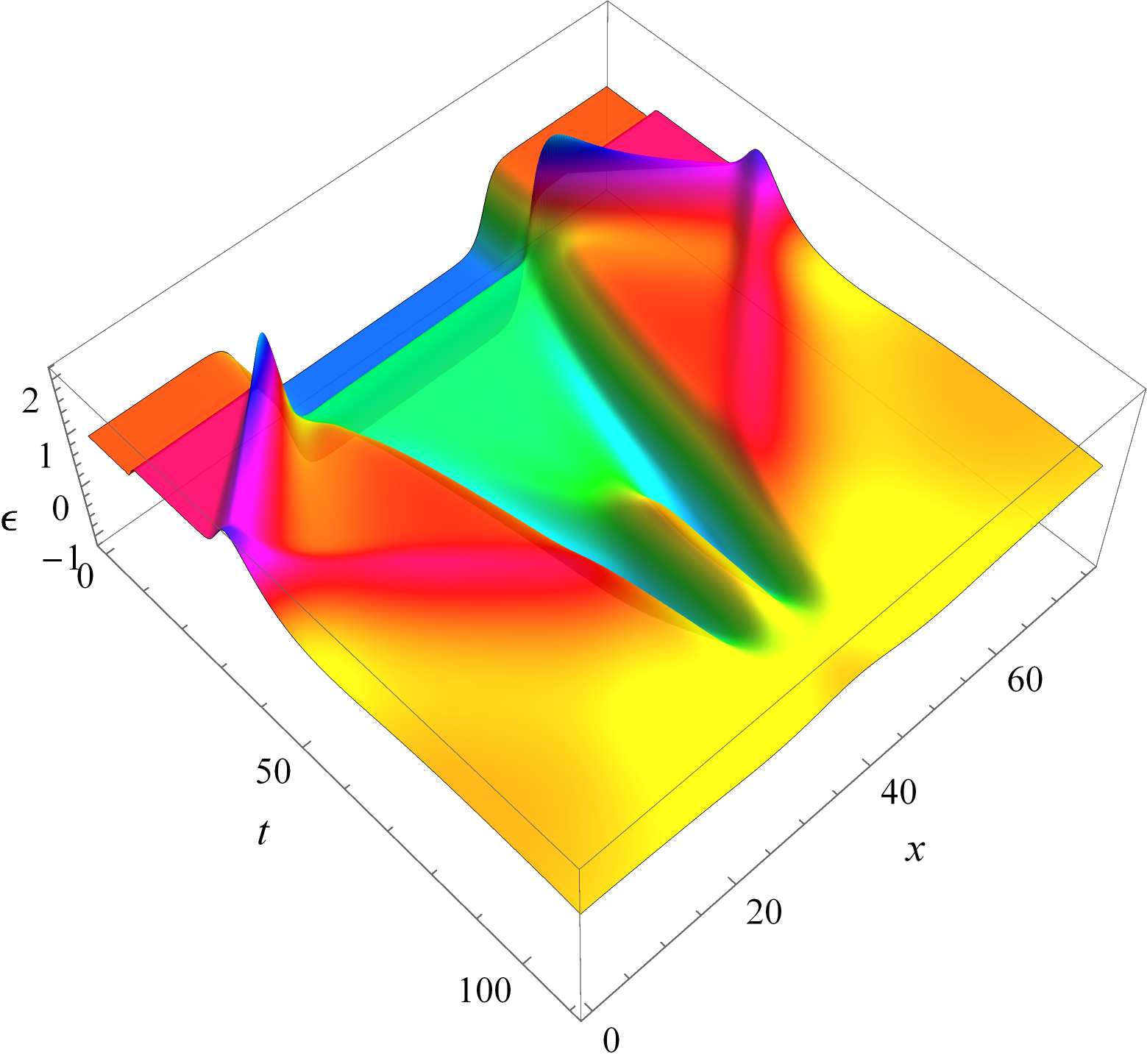}\label{fig:energy_density_homo}}
		\caption{(a): The average energy of the final state of the quench process as a function of quench strength. 
			When the quench strength is larger than the critical value $H_{*}$, the final state is a homogeneous state, on the contrary, it is still a phase-separated state.
			The final state of the quench process with strength $H_{A}<H$ lie on the upper branch with energy above that of point $A$ in Fig. \ref{fig:s-T}.
			That is to say, the final state is a supercooled state in the $AB$ region of Fig. \ref{fig:s-T} when the quench strength $H_{*}<H<H_{A}$.
			(b): The energy density as a function of time during the quench process with the {quench strength $H=0.44$} slightly greater than the critical quench strength $H_{*}$.}\label{fig:quench_H}
	\end{center}
\end{figure}

In order to further reveal the effect of quench strength on the system, we present the variation of the average energy of the final state with the quench strength in Fig. \ref{fig:energy_H}.
Our findings suggest that a stronger quench leads to more energy being added to the system.
Furthermore, a quench with sufficient strength (greater than the critical value $H_{*}$) can destroy the phase-separated structure and drive the system towards a homogeneous final state, as shown in Fig. \ref{fig:energy_density_homo}.
Interestingly, the final state obtained through a quench process with a strength $H_{*}<H<H_{A}$ is not the thermodynamically preferred state.
Instead, it is a metastable state located in the supercooled region exhibiting nonlinear instability. 
This implies that the system can undergo a dynamical transition to the phase-separated state again by introducing a seed nucleus, as observed in the process depicted in Fig. \ref{fig:ps_supercool_energy}.
However, when the quench strength exceeds $H_{A}$, the system is quenched to a stable homogeneous state lying on the upper branch with energy higher than that of point $A$ in Fig. \ref{fig:s-T}.
The results also indicate that although the energy of the system is always increased during such a quench process, the temperature of the final state may remain unchanged, decrease or increase, depending on the quench strength.

\subsection{Critical quench}
\begin{figure}[h!]
	\begin{center}
		\includegraphics[height=.26\textheight]{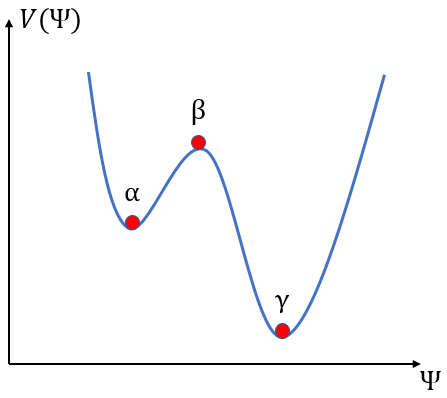}
		\caption{{The schematic diagram of the nonlinear instability, in which the horizonal and vertical axis represent the geometric configuration of the gravitational system and the related thermodynamic potential respectively. The local minimum points $\alpha$ and $\gamma$ represent the two local ground states of the system, connected by an excited state $\beta$.}}
		\label{fig:schematic}
	\end{center}
\end{figure}

{In the study of black hole dynamics, the linear dynamical instability has been widely revealed, which is manifested as the initial value possessing a dynamically unstable mode, such as superradiant instability \cite{Sanchis-Gual:2015lje,Bosch:2016vcp}, spontaneous scalarization \cite{Doneva:2017bvd,Silva:2017uqg,Antoniou:2017acq,Herdeiro:2018wub,Cunha:2019dwb,Herdeiro:2020wei,Berti:2020kgk} and spinodal instability \cite{Janik:2015iry,Janik:2017ykj}.
Under an arbitrarily small perturbation, the excited unstable mode will grow exponentially with time and push the system away from the initial state, leading to the occurrence of dynamical transition.
In this case, the linearly unstable initial state can be considered to be located on an excited state, which can spontaneously evolve to a ground state.
In addition, there exists a type of nonlinear dynamical instability, which is manifested in the fact that the occurrence of the corresponding dynamical transition requires the perturbation strength to exceed a non-zero threshold.
Such a threshold indicates the existence of a barrier in the dynamical transition.
Under this scenario, the initial state is dynamically stable under perturbations due to the lack of unstable modes, indicating that it resides on a local ground state.
On the other hand, the occurrence of the dynamical transition and the existence of the threshold indicate that there is another local ground state as the final state and an excited state as the barrier, as shown in Fig. \ref{fig:schematic}.
Under a sufficiently large disturbance, the system can dynamically transition from one local ground state to the other by crossing over the excited state.
When the disturbance strength approaches the threshold for dynamical transition, the system will stay in the excited state during the dynamical intermediate process.
Furthermore, this critical dynamical transition is generally bidirectional.
For example, through the accretion mechanism of the scalar field, a bald black hole and a scalarized black hole located in two local ground states can achieve bidirectional dynamical transition by crossing a critical scalarized black hole as the excited state \cite{Zhang:2021nnn,Zhang:2022cmu,Liu:2022fxy,Jiang:2023yyn,Chen:2023eru}.}

{In the holographic first-order phase transition, the critical dynamical transition shown in Fig. \ref{fig:ps_supercool_energy} indicates that the supercooled state and the phase-separated state are located in two local ground states, respectively.
When the disturbance strength approaches a threshold, the critical inhomogeneous state that appears during the evolution serves as a dynamical barrier in the transition process, which is located on the excited state.
Since the configuration of such a critical state appears as an energy well locally existing on the background of the homogeneous solution, the mechanism that triggers this dynamical transition is a inhomogeneous scalar field disturbance with a sufficiently large amplitude, which acts as a seed nucleus to induce the nucleation process.
Inspired by the bidirectional dynamical transition process of the bald and scalarized black holes}, the phase-separated state here is supposed to also suffer from nonlinear instability and is able to be driven to a supercooled state by some dynamical mechanism.
However, since the configuration of phase-separated state is far from that of critical state, the naive disturbance mechanism {(such as scalar field disturbances)} is incapable of implementing this dynamical process.
The expected dynamical mechanism is required to not only increase the energy of the system, but also evolve the configuration of the system to approach the critical state with global near-uniformity.
{This is different from the bidirectional transition between a bald black hole and a scalarized black hole, which can be achieved through the same scalar field disturbance mechanism.}

The holographic quench mechanism provides an effective approach to realize the dynamical transition from the phase-separated state to the supercooled state.
On the one hand, the quench mechanism is capable of injecting enough energy into the system.
As shown in Fig. \ref{fig:energy_H}, the energy of the quenching system increases monotonically with the increase of the quench strength.
Specifically, for a sufficiently large quench strength, the energy of the final state of the evolution exceeds the energy of the state represented by the point $B$ in Fig. \ref{fig:deltas-epsilon}, indicating that the system enters the energy region where the critical state exists.
Furthermore, the energy brought by the stronger quench process can even bring the system into the energy region with a single homogeneous solution.
On the other hand, the quench process strongly changes the configuration of the phase-separated state and gives it a tendency to evolve towards the configuration of the critical state.
As shown in Fig. \ref{fig:stage1} and Fig. \ref{fig:stage1_negative}, in the first stage of the quench process, the high-energy and low-energy phases are heated independently and the energy of the low-energy phase gradually approaches that of the high-energy phase due to the larger expectation value of the low-energy phase.
That is to say, the configuration of the system tends to be homogeneous after quenching.
At the end of this stage, an inhomogeneous energy flow is excited at the domain walls, opening the possibility for the formation of a critical nucleus. 

\begin{figure}
	\begin{center}
		\subfigure[]{\includegraphics[width=.49\linewidth]{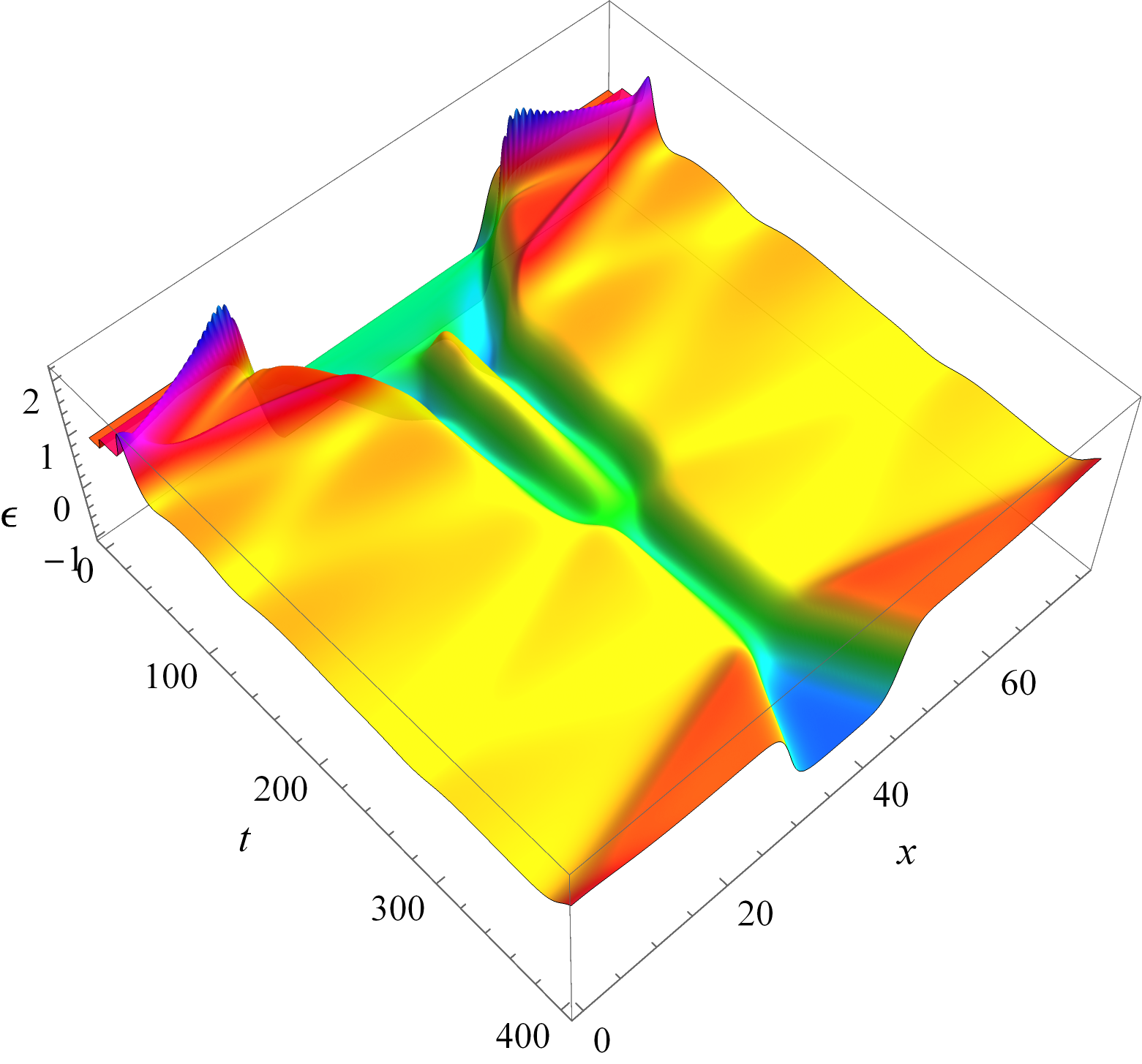}\label{fig:subcritical}}
		\subfigure[]{\includegraphics[width=.49\linewidth]{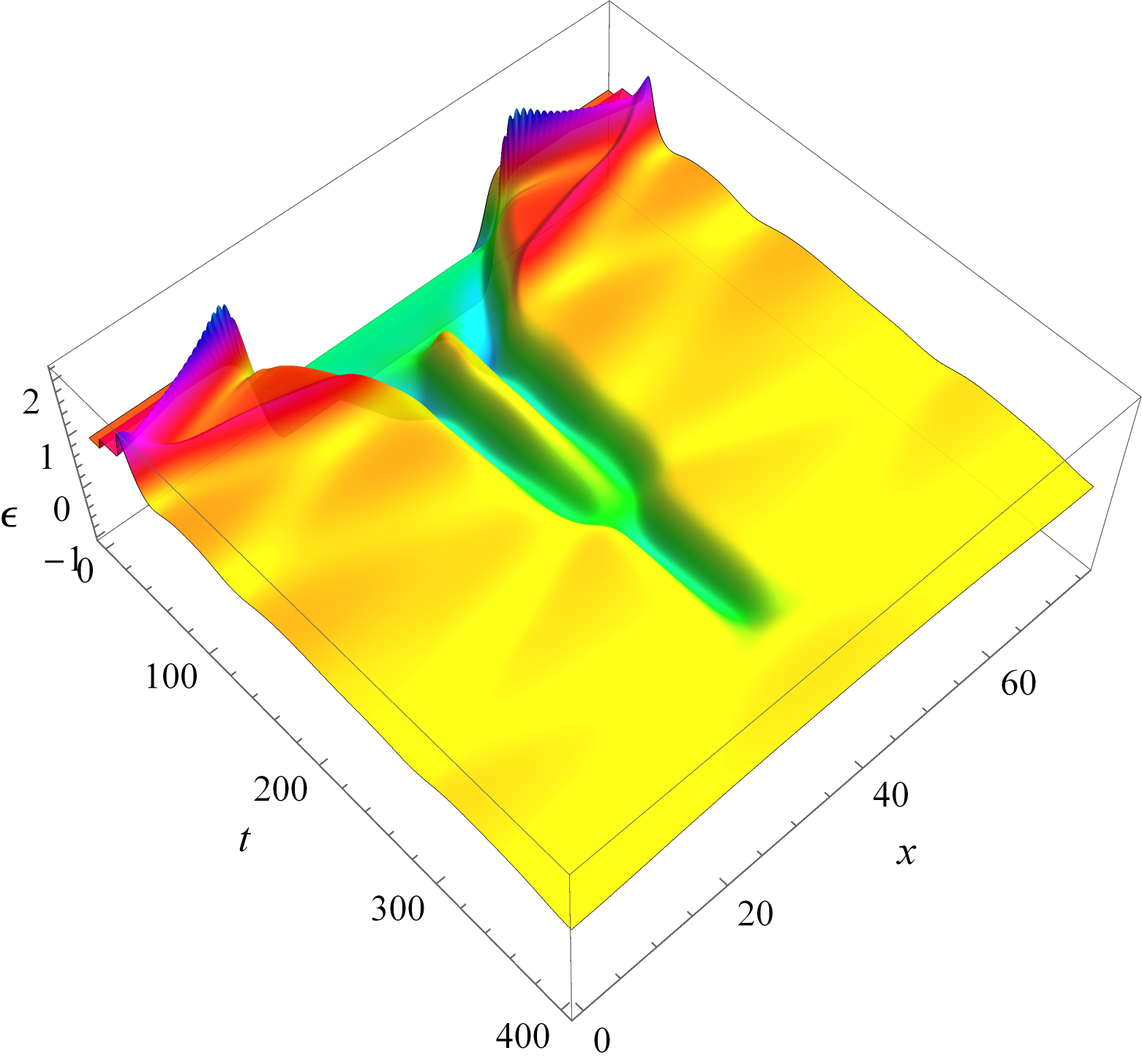}\label{fig:supercritical}}
		\caption{The real-time dynamics of the energy density for the case where the quench strength is near the critical value $H_{*}$. The left and right panels correspond to the cases of {subcritical strength ($H=0.42480111203125$) and supercritical strength ($H=0.424801112109375$)}, respectively.}\label{fig:critical_quench}
	\end{center}
\end{figure}

{From figure \ref{fig:schematic}, we expect that the transition from the phase-separated state to the supercooled state also requires crossing the critical nucleus as the excited state.
To this end, through the dichotomy method, we continuously approach the critical value $H_{*}$ of the quench strength in Fig. \ref{fig:energy_H} to reveal the critical dynamical behavior of the system.
The corresponding simulation results are shown in Fig. \ref{fig:critical_quench}, from which it can be observed that the system is attracted to a critical state of the critical nucleus configuration during the dynamical transition from the phase-separated state to the homogeneous supercooled state.}
Subsequently, for the case of subcritical parameter $H<H_{*}$, as shown in Fig. \ref{fig:subcritical}, the critical nucleus gradually grows up at the end of the evolution, forming a low-energy phase.
The formation of the low-energy phase indicates that the system returns to a phase-separated state with the same configuration as the initial state.
However, compared with the initial state, the space ratio occupied by the low-energy phase is greatly reduced due to the energy brought by the quench.
Instead, for the case of supercritical parameter $H>H_{*}$, as shown in Fig. \ref{fig:supercritical}, the critical nucleus gradually dissipates into the homogeneous background, leaving a globally homogeneous system.
Similar to the dynamical transition from the homogeneous supercooled state to the phase-separated state induced by the seed nucleus, the critical nucleus also acts as a dynamical barrier in the transition in the opposite direction caused by the quench. 
Such results are consistent with the critical scalarization phenomenon of black holes, where two states with nonlinear instability can be dynamically transformed into each other by crossing a linearly unstable critical state.
The difference lies in the dynamical mechanism.
Different from the phenomenon that the scalar field accretion mechanism can induce both scalarization and descalarization processes, the scalar field disturbance in holographic first-order phase transition is more suitable for the dynamical transition from a homogeneous state to a phase-separated state, while the transition in the opposite direction is competent by the heating process.

\begin{figure}
	\begin{center}
		\subfigure[]{\includegraphics[width=.49\linewidth]{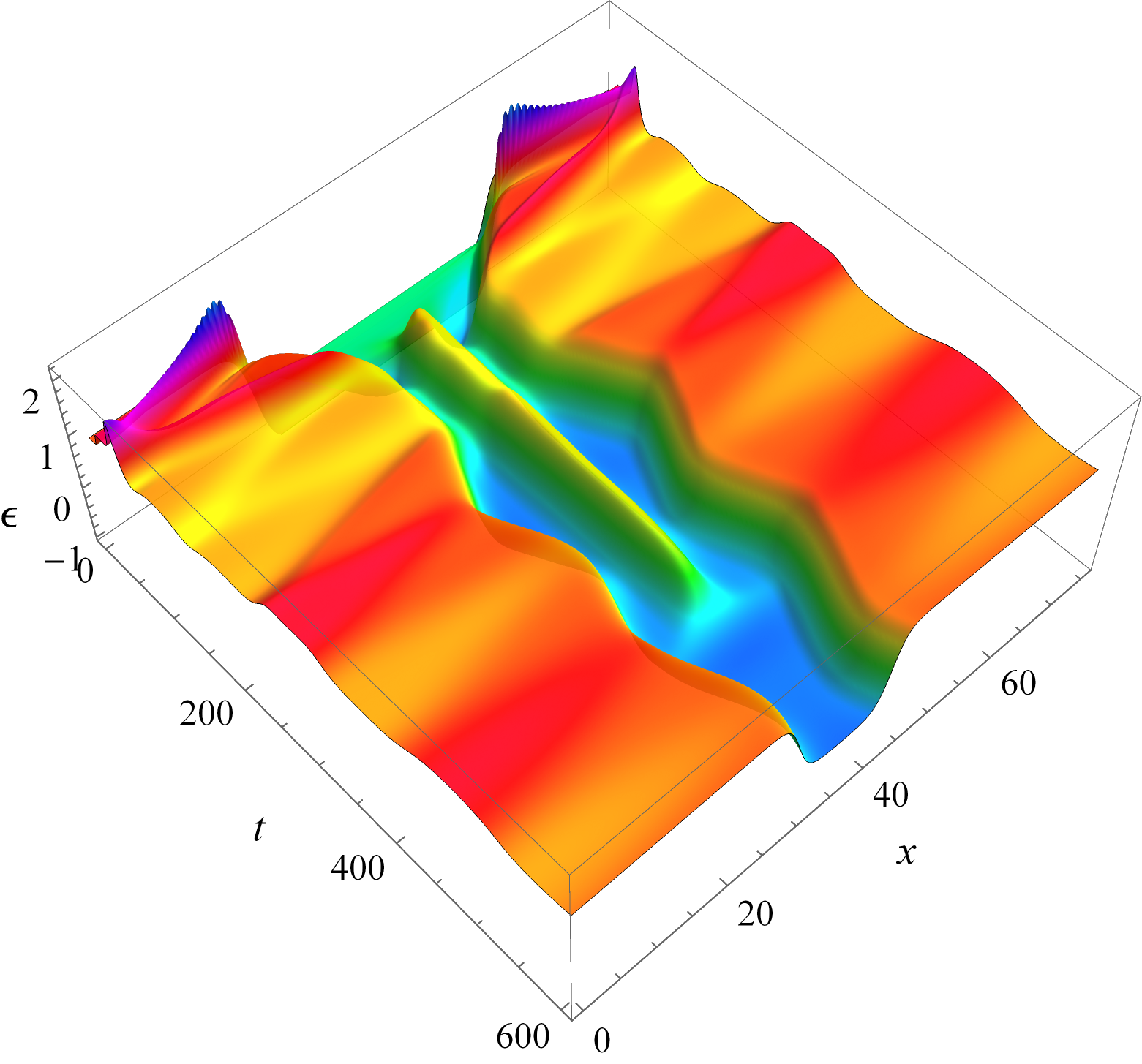}\label{fig:double}}
		\subfigure[]{\includegraphics[width=.49\linewidth]{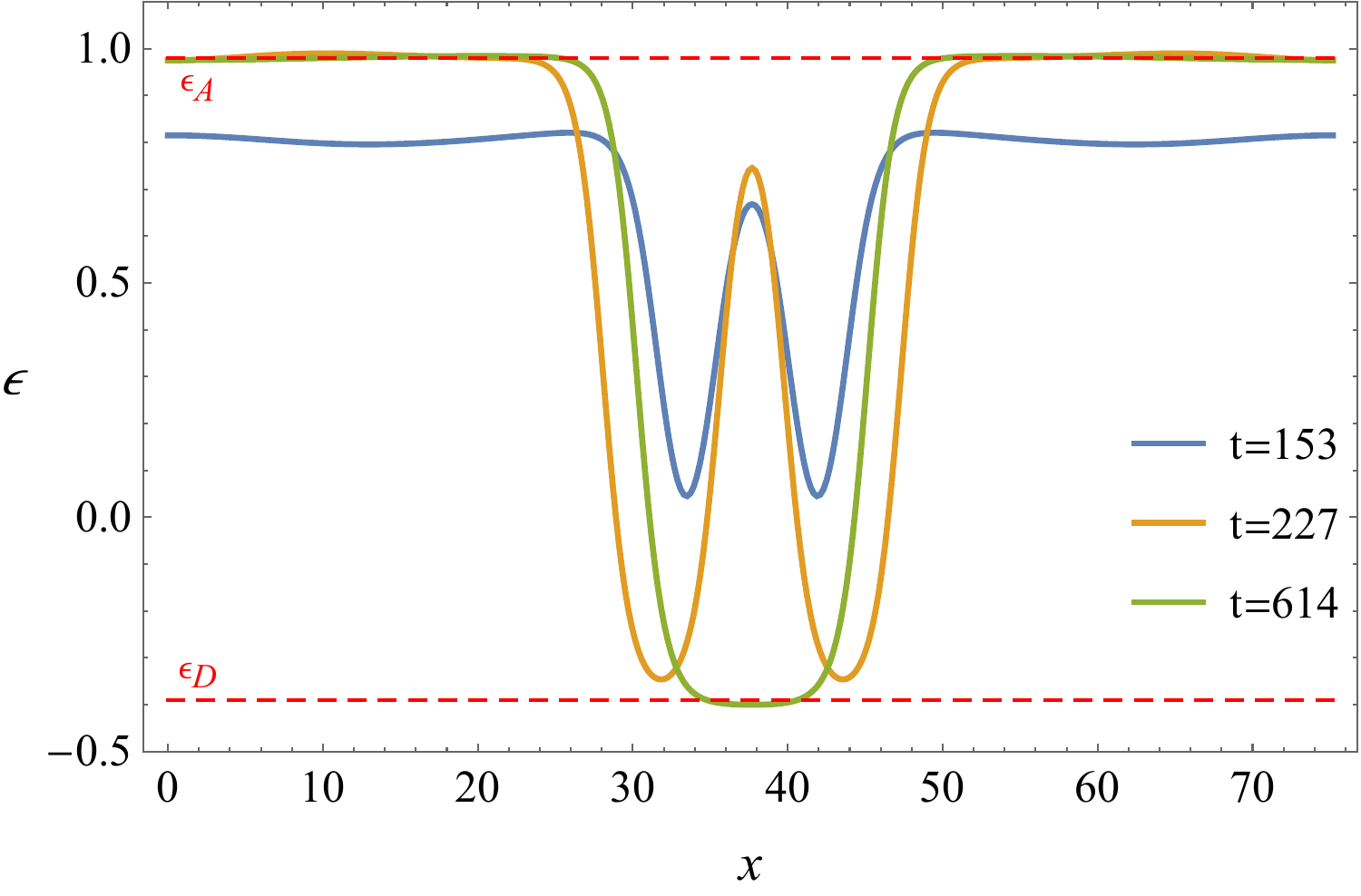}\label{fig:double_real}}
		\caption{(a): The temporal and spatial dependence of the energy density during the quench process with the {quench strength $H=0.424775$} slightly smaller than the critical value $H_{*}$.
			(b): The three characteristic energy density configurations in such dynamical process.}\label{fig:double_phase}
	\end{center}
\end{figure}

Interestingly, as seen in Fig. \ref{fig:critical_quench}, the critical nucleus configuration in such a dynamical process is formed by the fusion of double energy wells, which originate from the propagating energy flow from the domain walls to the low-energy phase caused by the quench.
In order to reveal the dynamical properties of such energy wells, we choose a quench strength slightly away from the critical value and show the real-time dynamics in Fig. \ref{fig:double_phase}.
Our numerical results show that each energy well is capable of nucleating individually and forming a low-energy region.
The system then briefly stays in a state with double low-energy regions connected by an energy peak.
However, such a configuration is dynamically linearly unstable.
Eventually, the energy peak gradually dissipates and the double low-energy regions fuse together to form a stable low-energy phase.
The three characteristic configurations experienced by the system during this dynamical process are shown in Fig. \ref{fig:double_real}.
This is the case where the quench strength is less than the critical value.
Conversely, for the quench strength greater than the critical value, as shown in Fig. \ref{fig:energy_density_homo}, the double energy wells each dissipate into the homogeneous background.
That is to say, the energy wells exhibit the dynamical properties of a critical nucleus: nucleation (for the subcritical quench strength) or dissipation (for the supercritical quench strength).
When the quench strength gradually approaches the critical value, the double energy wells merge together and converge to the linearly unstable critical nucleus, which further evolves into nucleation or dissipation.


\section{Conclusion}\label{sec:C}
In this paper, we have investigated the quench dynamics in the holographic model of first-order phase transition, utilizing the characteristic formalism.
Our study involved heating a phase-separated initial state and subsequently tracking its nonlinear dynamical evolution.

In the first step, we prepared the phase-separated initial state through various strategies. Specifically, for a low-energy initial state, we induced it by perturbing an unstable state in the spinodal region. Due to the linear instability, even the slightest perturbation will prompt the system to undergo phase separation. Conversely, for a high-energy initial state, we induced it by perturbing a metastable state that is subject to nonlinear instability. In such cases, the dynamical transition can still be triggered by applying a sufficiently strong disturbance that serves as a seed nucleus for the system.


In the second step, we investigated the dynamical behavior of the system in the phase-separated state under temporal dependence of the scalar source. Specifically, we introduced a thermal quench over a short period of time.
We have found that there are two distinct forms of response to the quenches with different signs.

Within this step, we further observed three characteristic stages in each response process.
The first stage can be viewed as a process in which the high-energy and low-energy phases are independently quenched since no energy flow has formed throughout the entire space. 
Furthermore, as indicated by the Ward-Takahashi identity (\ref{eq:energy_change}), the time-dependent scalar source influences the energy of the system via the expectation value of the scalar field.
Given the positive expectation value that the low-energy phase possesses, its energy variation over time is opposite to that of the scalar source. As a result, quenches of different signs will induce energy to vary in opposite ways over time.
Conversely, the energy of the high-energy phase hardly changes due to its nearly vanishing expectation value of the scalar field.
At the end of this stage, an intriguing phenomenon arises with the behavior of the domain walls, as they exhibit varying responses to the quenches with different signs.
Specifically, for the quench with a positive (negative) sign, two energy peaks (valleys) are excited at the domain walls.
In the second stage, these energy peaks (valleys) subsequently propagate deep into the high-energy phase region. Throughout this process, multiple groups of energy peaks or valleys are successively excited and absorbed, resulting in the formation of a large energy flow.
Ultimately, for a weak quench strength, as the energy flow gradually decays in space, the system settles down to a phase-separated state with higher energy in the third stage.

Furthermore, altering the quench strength typically results in a corresponding change in the final state. Specifically, as the quench strength increases, we observed a critical value of strength, denoted by $H_{*}$, that can break the phase-separated structure and dynamically drive the system to a homogeneous state.
As illustrated in the microcanonical phase diagram in Fig. \ref{fig:deltas-epsilon}, the homogeneous state with energy greater than the phase-separated state at point $b$ can take two forms. Specifically, it can be the metastable state located in the $AB$ region, or it can be the stable state with higher energy than the state at point $A$.
It's important to note that the system cannot remain in a state with an energy lower than that of point $B$ due to spinodal instability.
As illustrated in Fig. \ref{fig:energy_H}, our results show the existence of a second critical strength value, denoted by $H_A$. Specifically, when the quench strength lies within the range $H_{*}<H<H_{A}$, the system evolves into the metastable state below the phase transition temperature. Conversely, if $H_{A}<H$, the final state is a thermodynamically stable phase.

In the third step, we investigate the real-time dynamics of the system near the critical quench strength $H_{*}$. 
Such a quench can dynamically drive the system from a phase-separated state to a critical state with a critical nucleus configuration, which is dynamically unstable and evolves towards nucleation (for subcritical quench strengths) or dissipation (for supercritical quench strengths).
This dynamical property that different quench strengths lead to different evolution trends also occurs in local energy wells.
{The above results confirm that the homogeneous supercooled state and inhomogeneous phase-separated state are respectively located in two local ground states and can dynamically transform into each other by crossing the critical nucleus as an excited state.
Among them, scalar disturbance can trigger the transition from the supercooled state to the phase-separated state, while for the transition in the opposite direction, the holographic quench provides a more effective triggering mechanism.}

In the future, 
it is interesting to explore the effect of the topology on the boundary system. 
Due to the periodic boundary condition along $x$ direction, the system is located on a torus. 
The aperiodic boundary condition is introduced in AdS/BCFT duality \cite{Takayanagi:2011zk,Nozaki:2012qd}, or more recently, the double holography \cite{Almheiri:2019hni,almheiri2019entropy,Almheiri:2019yqk}. 
For the numerical setups in higher dimensions \cite{Almheiri:2019psy,Ling:2020laa,Ling:2021vxe,Liu:2022pan}, CFL conditions may have more stringent constraints on aperiodic boundary conditions.
Furthermore, to study the properties of strongly coupled quantum systems near phase transition points \cite{Fu:2022qtz,Fu:2022jqn,Liu:2020blk,Ling:2016wyr,Ling:2015dma}, one can observe the changes in entanglement entropy \cite{Ryu:2006bv,Casini:2011kv,Hubeny:2007xt,Ryu:2006ef} and complexity \cite{Brown:2015bva,Alishahiha:2015rta,Carmi:2017jqz,Brown:2015lvg} during the quench process \cite{Albash:2010mv,Liu:2013iza,Liu:2013qca,Chen:2018mcc,Chapman:2018dem,Chapman:2018lsv,Ling:2018xpc,Ling:2019ien,Ling:2019tbi,Bai:2014tla}. It would be also interesting to extend our analysis to some other gravity modes, such as the charged inhomogeneous system. 
In addition, it is meaningful to seek a hydrodynamic description for the quench process.

\appendix
\section{Numerical procedure for static solutions}\label{sec:Aa}
\subsection{The Einstein-DeTurck formalism}
The line element of the standard Schwarzschild-AdS$_4$ geometry is 
\begin{align}\label{eq_SAdS}
	ds^2&=\frac{1}{z^2}\left[-f(z)dt^2+\frac{dz^2}{f(z)}+dx^2+dx_1^2\right],\\
	f(z)&=1-
	\left(\frac{z}{z_h}\right)^3,
\end{align}
where $z=z_h$ is the locus of the bifurcation and the AdS radius $L$ has been set to $1$. For a regular metric on the bifurcation, the transformation of radial coordinate $z$ is imposed to be 
$$y=\sqrt{1-\frac{z}{z_h}},$$
which renders the conformal boundary to be $y=1$ and the locus of the bifurcation to be $y=0$.
With the break of the translational symmetry along $x$ direction, the most general metric reads
\begin{align}\label{eq_Background}
	ds^2=&\frac{1}{z_h^2(1-y^2)^2}\left[-y^2 P Q_1 dt^2 +\frac{4 z_h^2 Q_2}{P}dy^2+
	Q_4\left(dx + y (1-y)^3 Q_3 dy \right)^2+Q_5dx_1^2\right],\\
	\phi=&z_h(1-y^2)\,Q_6, \label{eq_Scalar}
\end{align}
with $P=3-3y^2+y^4$ and $Q_i\,(i=1,\cdots,6)$ being the functions of $(x,y)$.

Instead of solving Einstein equations (\ref{eq:fe}) directly, we solve the so-called Einstein-DeTurck equations, which are
\begin{align}\label{eq_EDE}
	R_{\mu\nu}+3 g_{\mu\nu}&=\left(T_{\mu\nu}-\frac{T}{2}g_{\mu\nu}\right)+\nabla_{(\mu}\xi_{\nu)},
\end{align}
where $$\xi^{\mu}:=\left[\Gamma_{\nu \sigma}^{\mu}(g)-\Gamma_{\nu \sigma}^{\mu}(\bar{g})\right] g^{\nu \sigma}$$ is the DeTurck vector and $\bar{g}$ is the reference metric, whose line element is chosen to be (\ref{eq_SAdS}). For Dirichlet boundaries, $\bar{g}$ is required to satisfy the same boundary conditions as $g$, but there is no requirement on Neumann boundaries
\cite{Almheiri:2019psy} (see also in \cite{Ling:2020laa,Liu:2022pan}).

The domain of interest is periodic in the $x$ direction, which is $x \in [0,2l]$. Considering the mirror symmetry at $x=l$, the whole domain of integration can be reduced to $x \in [0,l]$ together with $y \in [0,1]$. The full boundary conditions are shown in Tab.~\ref{tab_BCS}.

\begin{table}
	\centering
	\begin{tabular}{|c| c| c |c |c| c |c|}
		\hline
		& 1 & 2 & 3 & 4 & 5 & 6 \\
		\hline
		$\mathbf{y=1}$ & $Q_1=1$ & $Q_2=1$ & $Q_3=0$ & $Q_4=1$ & $Q_5=1$ & $Q_6=1$\\
		\hline
		$\mathbf{y=0}$ & $\partial_y Q_1=0$ & $\partial_y Q_2=0$ & $\partial_y Q_3=0$ & $\partial_y Q_4=0$ & $\partial_y Q_5=0$ & $\partial_y Q_6=0$\\
		\hline
		$\mathbf{x=l}$ & $\partial_x Q_1 =0$ & $\partial_x Q_2 =0$ & $Q_3=0$ & $\partial_x Q_4 =0$  & $\partial_x Q_5 =0$ & $\partial_x Q_6 =0$\\
		\hline
		$\mathbf{x=0}$ & $\partial_x Q_1 =0$ & $\partial_x Q_2 =0$ & $Q_3=0$ & $\partial_x Q_4 =0$  & $\partial_x Q_5 =0$ & $\partial_x Q_6 =0$\\
		\hline
	\end{tabular}
	\caption{Boundary conditions.} \label{tab_BCS}
\end{table}

Due to the non-periodicity, both directions are discretized by the Chebyshev Pseudo-spectral method with Gauss–Lobatto grids. Subsequently, the geometry can be solved numerically by the Newton-Raphson iteration.

The boundary conditions at the
bifurcation $y=0$ will further imply that $Q_1(x,0)=Q_2(x,0)$, which
fixes the Hawking temperature of the black hole as
\begin{equation}\label{eq_HawkingTemp4}
	T_h=\frac{3}{4\pi z_h}. 
\end{equation}
\subsection{Asymptotic expansions}
In the Fefferman-Graham expansion of $\{Z,\chi\}$, the asymptotic behaviors of $Q_i$ near the conformal boundary $Z=0$ are obtained as
\begin{align}
	Q_1&=1-\frac{Z^2}{8}-\frac{Z^3\left(3 z_h^2+Q_1^{(0,3)}[\chi, 1]\right)}{48 z_h^3}+\cdots,\\
	Q_2&=1-\frac{Z^3 Q_6^{(0,1)}[\chi, 1]}{6 z_h}+\cdots,\\
	Q_3&=-\frac{Z Q_3^{(0,1)}[\chi, 1]}{2 z_h}+\frac{Z^2\left(-Q_3^{(0,1)}[\chi, 1]+Q_3^{(0,2)}[\chi, 1]\right)}{8 z_h^2}+\cdots,\\
	Q_4&=1-\frac{Z^2}{8}-\frac{Z^3\left(3 z_h^2+Q 4^{(0,3)}[\chi, 1]\right)}{48 z_h^3}+\cdots,\\
	Q_5&=1-\frac{Z^2}{8}+\frac{Z^3\left(6 z_h^2+24 z_h^2 Q_6^{(0,1)}[\chi, 1]+Q_1^{(0,3)}[\chi, 1]+Q_4^{(0,3)}[\chi, 1]\right)}{48 z_h^3}+\cdots,\\
	Q_6&=1-\frac{Z Q_6^{(0,1)}[\chi, 1]}{2 z_h}-\frac{193 Z^2}{720}+\cdots,
\end{align}
where the derivatives $Q_i^{(0,j)}\, (i=\{1,\cdots,6\};\, j=\{1,2,3\})$ are related to coordinate $y$. For instance, $Q_6^{(0,1)}[\chi, 1]$ means that $Q_6^{(0,1)}[\chi, y]\bigg|_{y=1}$.

With these expansions in hand, the entropy and energy density can be expressed by
\begin{align}
	s(\chi)&=\frac{2\pi}{z_h^2}\sqrt{Q_4[\chi,0]Q_5[\chi,0]},\\
	e(\chi)&=\frac{96+9 z_h^2 +32 z_h^2 Q_6^{(0,1)}[\chi, 1]+3 Q_1^{(0,3)}[\chi, 1]}{96 z_h^3},
\end{align}
respectively. 
Hence, the averaged entropy and energy can be further integrated as
\begin{align}
	\bar{S}&=\frac{1}{l}\int_0^ls(\chi) d\chi,\\
	\bar{E}&=\frac{1}{l}\int_0^le(\chi) d\chi.
\end{align}

\section{Numerical procedure for dynamic evolution}\label{sec:Ab}
\subsection{Evolution algorithm}
For the nonlinear evolution, the coupled field equations (\ref{eq:fe}) are solved numerically using the characteristic formulation method in \cite{Chesler:2013lia}, which has been shown to be applicable to various gravitational dynamics problems in the asymptotically AdS spacetime.
The ansatz for the metric is given by
\begin{equation}
	ds^{2}=\Sigma^{2}\left(Gdx^{2}+G^{-1}dy^{2}\right)+2dt\left(dr-Adt-Fdx\right),
\end{equation}
where all fields are functions of $(t,x,r)$.
To simplify the problem, we have assumed that the system is translation invariance in the $y$ direction. 
Note that such a form of ansatz is invariance under the radial shifts
\begin{subequations}
	\begin{align}
		r&\rightarrow\bar{r}=r+\lambda(t,x),\\
		A&\rightarrow\bar{A}=A+\partial_{t}\lambda(t,x),\\
		F&\rightarrow\bar{F}=F+\partial_{x}\lambda(t,x),
	\end{align}
\end{subequations}
and in general, there are two ways to fix the parameter $\lambda$.
One simple way is to require $\lambda$ to disappear throughout the evolution.
However, in our work, we use the reparameterization freedom to put the apparent horizon at a fixed radial position $r=1$.
In this case, since the computational domain is a fixed interval, we can conveniently discretize the fields with pseudospectral methods.
The Chebyshev and Fourier pseudospectral are used to discretize the fields in the radial $(z)$ and spatial $(x)$ direction, respectively.
Note that we compactify the radial coordinate by $z=r^{-1}$ such that the AdS boundary is at the position of $z=0$.

In order to decouple the field equations, we introduce a derivative operator $d_{+}=\partial_{t}-z^{2}A\partial_{z}$, which is the directional derivative along the outgoing null geodesic.
At this point, the set of equations (\ref{eq:fe}) has the following nested structure
\begin{subequations}
	\begin{align}
		&\Sigma'' + \frac{2 }{z}\Sigma'+\frac{1}{4}\left( \frac{ G'^{2}}{G^{2}} +\phi'^{2}\right)\Sigma=0,\label{eq:Sigma}\\
		&F''+\left( \frac{2}{z}-\frac{G'}{G}\right)F'+\left(\partial_{z}+\frac{2}{z}-\frac{G' }{G} \right)\left(\frac{2  \Sigma'}{ \Sigma}-\frac{G'}{G} \right) F=S_{F}\left[\phi,G,\Sigma\right],\label{eq:F}\\
		&(d_{+}\Sigma)'+\frac{\Sigma'}{\Sigma}d_{+}\Sigma=S_{d_{+}\Sigma}[\phi,G,\Sigma,F],\label{eq:Sigma_plus}\\
		&(d_{+}G)'+\left(\frac{\Sigma'}{\Sigma}-\frac{G'}{G}\right)d_{+}G=S_{d_{+}G}[\phi,G,\Sigma,F,d_{+}\Sigma],\label{eq:G_plus}\\
		&(d_{+}\phi)'+\frac{\Sigma'}{\Sigma}d_{+}\phi=S_{d_{+}\phi}[\phi,G,\Sigma,F,d_{+}\Sigma],\label{eq:phi_plus}\\
		&A''+\frac{2}{z}A'=S_{A}[\phi,G,\Sigma,F,d_{+}\Sigma,d_{+}G,d_{+}\phi],\label{eq:A}\\
		&(d_{+}F)'-\left(\frac{2\Sigma'}{\Sigma}+\frac{G'}{G}\right)d_{+}F=S_{d_{+}F}[\phi,G,\Sigma,F,d_{+}\Sigma,d_{+}G,d_{+}\phi,A],\label{eq:F_plus}\\
		&d_{+}d_{+}\Sigma=S_{d_{+}d_{+}\Sigma}[\phi,G,\Sigma,F,d_{+}\Sigma,d_{+}G,d_{+}\phi,A,d_{+}F],\label{eq:Sigma_plus2}
	\end{align}\label{eq:cf}
\end{subequations}
where the prime stands for the derivative with respect to the variable $z$.
Once given the data for the fields $\phi$ and $G$ on the time slice $t_{0}$, the above equations can be solved sequentially, since the source terms given on the right-hand side of the equations depend only on the known fields.
In fact, the last two equations are not solved but are used to detect numerical errors.
The reason this can be done is that once the field $A$ is solved from the equation (\ref{eq:A}), the fields $\phi$ and $G$ can be pushed to the next time slice $t_{0}+dt$ by integrating in time for $\partial_{t}\phi$ and $\partial_{t}G$, which can be obtained from the auxiliary fields
\begin{subequations}
	\begin{align}
		\partial_{t}\phi&=d_{+}\phi+z^{2}A\phi',\\
		\partial_{t}G&=d_{+}G+z^{2}AG'.
	\end{align}
\end{subequations}
The procedure is iterated until the entire simulation is completed.

\subsection{Boundary conditions}
In order to clarify the boundary conditions, we expand the field equations (\ref{eq:cf}) near the AdS boundary and obtain the asymptotic behavior of the fields as follows
\begin{subequations}
	\begin{align}
		\phi&=\phi_{1}z+\phi_{2}z^{2}+o(z^{3}),\label{eq:asy_phi}\\
		G&=1+g_{3}z^{3}+o(z^{4}),\label{eq:asy_G}\\
		\Sigma&=z^{-1}+\lambda-\frac{\phi^{2}_{1}}{8}z+o(z^{2}),\label{eq:asy_Sigma}\\
		F&=-\partial_{x}\lambda+f_{1}z+o(z^{2}),\label{eq:asy_F}\\
		A&=\frac{1}{2}z^{-2}+\lambda z^{-1}+\frac{1}{2}\lambda^{2}-\frac{\phi^{2}_{1}}{8}-\partial_{t}\lambda+a_{1}z+o(z^{2}),\label{eq:asy_A}
	\end{align}
\end{subequations}
where the source of the scalar field $\phi_{1}$ is a free parameter for the field equations, which is used to quench the system in our work.
Since the asymptotic behavior of the fields is known, the scalar operator and energy-momentum tensor operator can be evaluated immediately from the equations (\ref{eq:operators})
\begin{subequations}
	\begin{align}
		\left\langle O\right\rangle &=\frac{1}{2}\left(\lambda\phi_{1}+\phi_{2}-\partial_{t}\phi_{1}\right),\\
		\left\langle T_{ij}\right\rangle&=\begin{pmatrix}
			-2a_{1}-\phi_{1}\left\langle O\right\rangle&-\frac{3}{2}f_{1}+\frac{1}{4}\phi_{1}\partial_{x}\phi_{1}&0\\
			-\frac{3}{2}f_{1}+\frac{1}{4}\phi_{1}\partial_{x}\phi_{1}&-a_{1}+\frac{3}{2}g_{3}&0\\
			0&0&-a_{1}-\frac{3}{2}g_{3}
		\end{pmatrix}.
	\end{align}
\end{subequations}
Meanwhile, the Ward-Takahashi identity (\ref{eq:WD}) gives rise to the time derivatives of $a_{1}$ and $f_{1}$ as follows
\begin{subequations}
	\begin{align}
		\partial_{t}a_{1}&=\frac{3}{4}\partial_{x}f_{1}-\frac{1}{2}\phi_{1}\partial_{t}\left\langle O\right\rangle-\frac{1}{8}\left(\partial_{x}\phi_{1}\right)^{2}-\frac{1}{8}\phi_{1}\partial^{2}_{x}\phi_{1},\label{eq:pt_a}\\
		\partial_{t}f_{1}&=\frac{2}{3}\partial_{x}a_{1}-\partial_{x}g_{3}+\frac{2}{3}\partial_{x}\phi_{1}\left\langle O\right\rangle+\frac{1}{6}\partial_{t}\phi_{1}\partial_{x}\phi_{1}+\frac{1}{6}\phi_{1}\partial_{x}\partial_{t}\phi_{1}.\label{eq:pt_f}
	\end{align}
\end{subequations}
Actually, the above equations can also be obtained by asymptotic analysis of equations (\ref{eq:F_plus}) and (\ref{eq:Sigma_plus2}) on the boundary, respectively.

With these preliminaries in hand, we can now start discussing how to solve these equations (\ref{eq:cf}).
First, the equation (\ref{eq:Sigma}) is a linear second-order ordinary differential equation for the field $\Sigma$.
The two integration constants are fixed by the coefficients of the first and second terms in the asymptotic behavior (\ref{eq:asy_Sigma}), which indicates the parameter $\lambda$ must be considered as the dynamical variable like $\phi$ and $G$.
The corresponding time derivative term can be extracted from the asymptotic behavior of field $A$ (\ref{eq:asy_A})
\begin{equation}
	\partial_{t}\lambda=\lim\limits_{z\rightarrow 0}\left(\frac{1}{2}z^{-2}+\lambda z^{-1}+\frac{1}{2}\lambda^{2}-\frac{\phi^{2}_{1}}{8}-A\right).
\end{equation}
Second, the coefficient of the subleading term $f_{1}$ in the asymptotic behavior of field $F$ (\ref{eq:asy_F}) provides the single integration constant for the equation (\ref{eq:F}) due to the vanishing coefficient of the $z^{-2}$ term.
The value of its time derivative can be easily calculated from equation (\ref{eq:pt_f}).
Then, the apparent horizon condition is used to fix the single needed integration constant for the equation (\ref{eq:Sigma_plus})
\begin{equation}
	{d_{+}\Sigma}(z_{h},t,x)=\left.\frac{1}{2G\Sigma}\left(\frac{z^{2} {F}^{2}  \Sigma'}{\Sigma} + \frac{{F} \partial_{x} G}{G}-\partial_{x} {F}\right)\right|_{z_{h}}.\label{eq:HP}
\end{equation}
Next, for the equations (\ref{eq:G_plus}) and (\ref{eq:phi_plus}), the integration constants are fixed by the coefficients of the first term in the asymptotic behavior $d_{+}G\sim -\frac{3}{2}g_{3}z^{2}+o(z^{3})$ and the second term in the asymptotic behavior $d_{+}\phi\sim-\frac{1}{2}\phi_{1}-\left(\lambda \phi_{1}+\phi_{2}-\partial_{t}\phi_{1}\right)z+o(z^{2})$, respectively.
As we said before, we require the position of the apparent horizon to be time-invariant, meaning that the time derivative of the condition (\ref{eq:HP}) is also always satisfied at the apparent horizon.
Using the equation (\ref{eq:Sigma_plus2}), the stationary horizon condition can induce a second-order elliptic equation for $A$ at the horizon, which fixes one of the two integration constants for equation (\ref{eq:A}).
The other is fixed by the coefficient of the leading term in the asymptotic behavior (\ref{eq:asy_A}).
The numerical code is available at \href{https://github.com/QianChen2022/HFOPT}{https://github.com/QianChen2022/HFOPT}.

\subsection{Convergence test}
\begin{figure}
	\begin{center}
		\subfigure[]{\includegraphics[width=.49\linewidth]{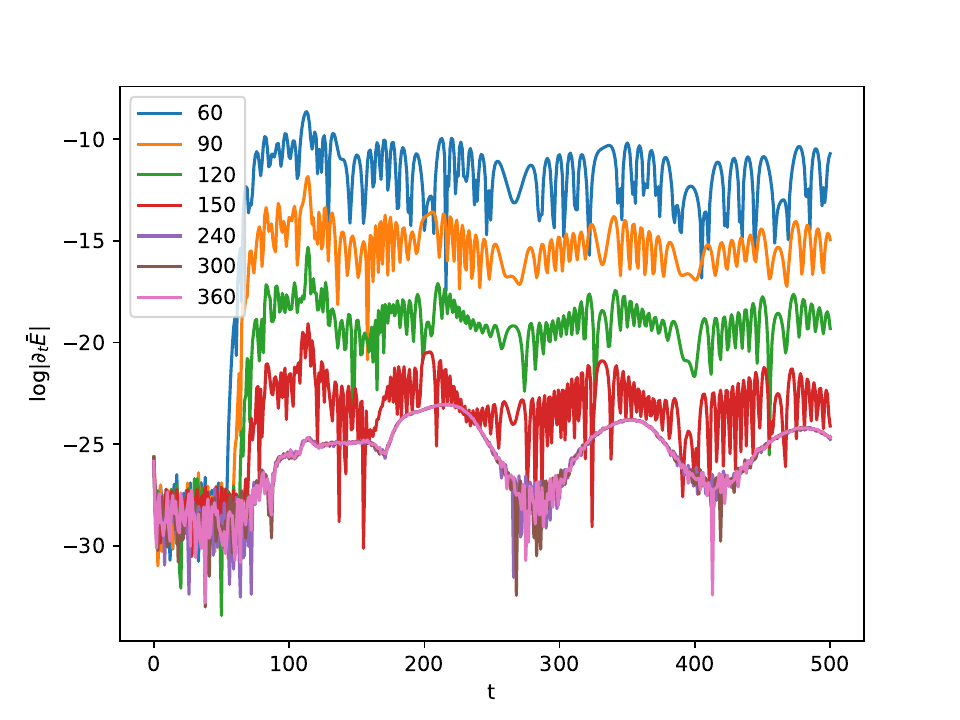}\label{fig:accuracy}}
		\subfigure[]{\includegraphics[width=.49\linewidth]{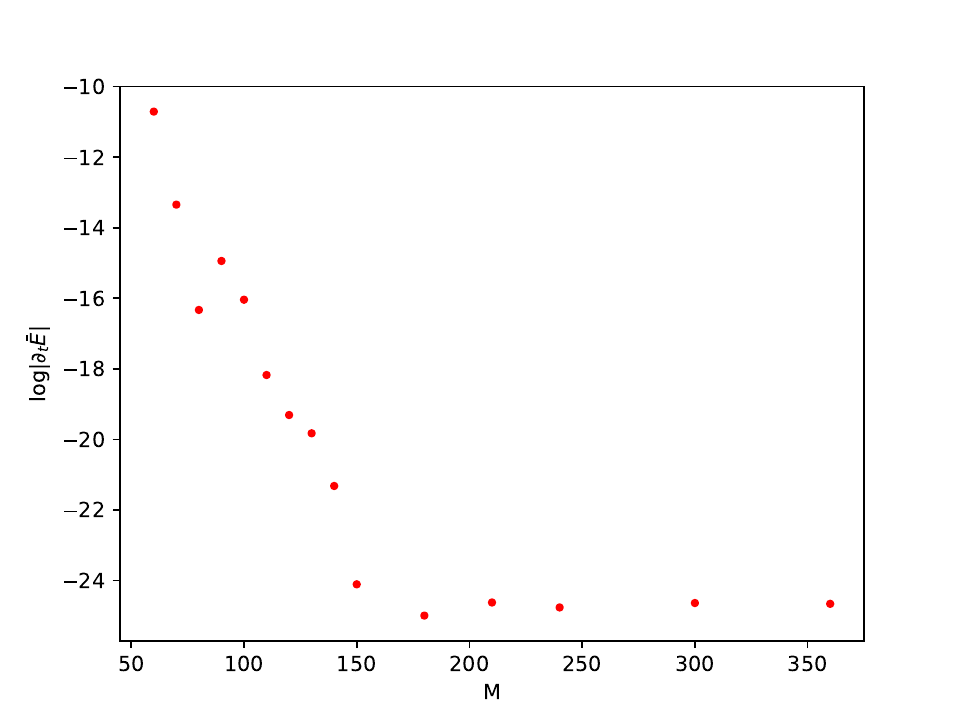}\label{fig:convergence}}
		\caption{The accuracy and convergence of our numerical code during the spontaneous dynamical transition in Fig. \ref{fig:ps_spinodal_energy}.
			(a): The value of $\text{log}|\partial_{t}\bar{E}|$ as a function of time with different numbers of grid points along the spatial direction.
			(b): The value of $\text{log}|\partial_{t}\bar{E}|$ at $t=500$ varies with the numbers of grid points along the spatial direction.}\label{fig:convergence_test}
	\end{center}
\end{figure}

In our integration strategy, the redundant boundary condition (\ref{eq:pt_a}) induced by equation (\ref{eq:Sigma_plus2}) on the AdS boundary allows us to detect numerical errors during the evolution.
Without loss of generality, we take the case of spontaneous dynamical transition in Fig. \ref{fig:ps_spinodal_energy} as an example to demonstrate the accuracy and convergence of our numerical code.
At this time, the condition (\ref{eq:pt_a}) is exactly the energy conservation condition $\partial_{t}\bar{E}=0$ due to the time-independent scalar source.
Since the homogeneous state as the initial data is obtained by the Newton-Raphson iteration algorithm, in order to avoid the influence of the iterative error, we keep the number of grid points along the radial direction $(z)$ and vary the number of grid points along the spatial direction $(x)$.
Due to the spectral method, the expected convergence rate is exponential.
The results in Fig. \ref{fig:convergence_test} show that the accuracy of our numerical code improves exponentially as the grid points increase, which is exactly the spectral accuracy.

\section*{Acknowledgement}
This research is partly supported by the National Key Research and Development Program of China Grant No. 2021YFC2203001 and the Natural Science Foundation of China (NSFC) under Grant Nos. 11731001, 11975235, 12035016, 12075026, 12275350 as well as by China Postdoctoral Science Foundation, under the National Postdoctoral Program for Innovative Talents BX2021303.

\bibliographystyle{aps}
\bibliography{references}

\end{document}